\documentstyle[12pt,epsf]{article}

\newcommand{\beq}{\begin{equation}}
\newcommand{\eeq}{\end{equation}}
\newcommand{\beqa}{\begin{eqnarray}}
\newcommand{\eeqa}{\end{eqnarray}}
\newcommand{\no}{\nonumber}
\newcommand{\q}{\quad}
\newcommand{\qq}{\qquad}
\newcommand{\mnod}{\stackrel{\circ}{m}}
\newcommand{\Fnod}{\stackrel{\circ}{F}}
\newcommand{\tr}{\mbox{tr}}
\newcommand{\ci}{\mbox{i}}
\newcommand{\e}{\mbox{e}}
\begin{document}

\hfill 

\hfill 

\bigskip\bigskip

\begin{center}

{{\Large\bf Non-leptonic Hyperon Decays\\ in Chiral Perturbation Theory
     \footnote{Work supported in part by the Deutsche Forschungsgemeinschaft
               and by the National Science Foundation}}}

\end{center}

\vspace{.4in}

\begin{center}
{\large B. Borasoy\footnote{email: borasoy@het.phast.umass.edu}
 and Barry R. Holstein\footnote{email: holstein@het.phast.umass.edu}}

\bigskip

\bigskip

Department of Physics and Astronomy\\
University of Massachusetts\\
Amherst, MA 01003\\

\vspace{.2in}

\end{center}

\vspace{.7in}

\thispagestyle{empty} 

\begin{abstract}
The non-leptonic hyperon decays are analyzed 
up to one-loop order
including {\it all} counterterms in the framework of heavy baryon
chiral perturbation theory. We use the exchange of the spin-$\frac{3}{2}$
decuplet resonances as an indication of which low-energy constants
contribute significantly to these investigated processes .
We choose four independent
decay amplitudes that are not related by isospin relations
in order to perform a fit for the pertinent low-energy constants 
and find a satisfactory fit both for s- and p-waves.
The chiral corrections to the lowest order forms
for the s-waves are moderate whereas there are significant
modifications of the p-wave amplitudes.
\end{abstract}

\vfill

\section{Introduction}
For three decades non-leptonic hyperon decays have been examined 
using effective field theories \cite{DGH}. There exist seven such transitions: 
$\Sigma^+  \rightarrow n \, \pi^+ \, , \, 
\Sigma^+  \rightarrow p \, \pi^0 \, , \, 
\Sigma^- \rightarrow n \, \pi^- \, , \, 
\Lambda \rightarrow p \, \pi^- \, ,  \,
\Lambda \rightarrow n \, \pi^0 \, , \,
\Xi^- \rightarrow  \Lambda \, \pi^-\,\mbox{and} \, 
\Xi^0 \rightarrow  \Lambda \, \pi^0 $, and
the matrix elements of these decays can be expressed in terms of 
parity-violating and parity-conserving amplitudes---s- and p-waves, 
respectively.
The weak $\Delta S = 1$ Hamiltonian transforms under $SU(3)\times SU(3)$
as $(8_L,1_R) \oplus (27_L,1_R)$ and, experimentally, the octet piece dominates
by a factor of twenty or so.
Therefore, we shall neglect the 27-plet contribution in what follows.

Chiral perturbation theory is
a systematic expansion in terms of small four-momenta $p$ and the current
masses $m_q$ of the light quarks, $q= u,d,s$.
In the case of non-leptonic hyperon decays lowest order 
chiral perturbation theory makes definite predictions for the
decay amplitudes in terms of just two weak couplings---the familiar
$f , d$ terms which parametrize the coupling of the octet weak spurion to
$\bar{B}^{\prime} B$ pairs.
These terms are examples of so-called low-energy constants (LECs),
{\it i.e.} coupling constants not fixed by chiral symmetry.
It has long been known that
if one employs values for the LECs which provide a good fit to the s-waves
then a poor fit is given for the p-waves.
On the other hand, a good p-wave representation yields a poor s-wave fit
\cite{DoGoHo}.
In the paper of Bijnens et al. \cite{Bij}, a first attempt was made in 
calculating the leading chiral corrections to these decays. However, 
the resulting
s-wave predictions no longer agreed with the data, and for the p-waves
corrections were even larger.

More recently, Jenkins reinvestigated this topic within the heavy baryon
formulation, including the spin-$\frac{3}{2}$ decuplet in the
effective theory \cite{Jen}.
But as in  \cite{Bij}
no counterterms were included---only the leading non-analytic,
{\it i.e.} ``leading log'', pieces from the loops were 
retained and $m_u = m_d = 0$ was assumed.
She found large cancellations between the octet and decuplet pieces in the 
loops, and therefore that the overall leading logarithmic
chiral correction is reduced. 
For the s-waves,
good agreement between theory and experiment was restored.
However, in the case of the p-waves, 
the chiral corrections did not lead to a satisfactory
description of the data. Indeed the lowest order p-wave
contribution consists of 
two baryon pole terms which tend to cancel to a large extent, 
enhancing the loop
corrections. In the p-waves then, one finds significant 
$SU(3)$ violation but not
necessarily a breakdown of the chiral expansion.
In order to obtain better understanding for the p-waves 
one should account for {\it all} terms
at one loop order, not just the leading log corrections
and that is the goal of our work.

This paper is organized as follows. In sec.~2 we write down the 
effective meson-baryon Lagrangian necessary to investigate the
non-leptonic hyperon decays, and spin-$\frac{3}{2}$ decuplet
resonance exchange is used as an indication of which terms of this
Lagrangian contribute significantly  to the decay amplitudes. 
There remain ten terms. Four of these higher order terms can be absorbed by
the lowest order terms since they amount to quark mass renormalizations of
the latter. 
We are not able to get a satisfactory fit for the decay amplitudes
by neglecting all other LECs in the Lagrangian. 
In \cite{Neu} a rough estimate of the LECs of the weak baryon Lagrangian
of order ${\cal O}(p)$ has been given using the weak deformation model
which lead to significant contributions to the p-waves. We will take these
LECs into account leading to a total number of ten coupling constants. 
In sec.~3 the theoretical calculation of the decay amplitudes 
is presented. Sec.~4 deals with the comparison of this computation
with experiment. A least-squares fit for the parameters is performed 
for the case of 
estimating the LECs solely by means of resonance saturation
yielding a very unsatisfactory fit both for s- and p-waves.
However, assuming non--vanishing counterterms of order ${\cal O}(p)$
in the p-wave sector, 
we achieve a much better fit for the decay amplitudes.
There remain, however, significant higher order corrections for the p-waves.
A short summary is given in sec.~5.
The complete effective Lagrangian, some technicalities and the
$Z$-factors are relegated to the appendices.

\section{Effective Lagrangian}
We perform our calculations using an effective Lagrangian within 
the heavy baryon
formalism. To this end, one writes down the most general relativistic
Lagrangian which is invariant under chiral and $CPS$ transformations,
the construction principles of which are outlined in app.~\ref{app.a}.
Imposing invariance of the Lagrangian under the transformation $S$
which interchanges down and strange quarks in the Lagrangian
one can further reduce the number of counterterms.
We will work in the $CP$ limit so that all LECs are real.
This Lagrangian is then reduced to the heavy fermion limit
by the use of path integral methods, which deliver
the relativistic corrections 
as $1/ \!\! \mnod$ terms in higher orders. 
The baryons are described by a
four-velocity $v_{\mu}$ and a consistent chiral counting scheme emerges,
{\it i.e.} a one-to-one correspondence between the Goldstone boson loops
and the expansion in small momenta and quark masses.
However, we will not present here the relativistic Lagrangian explicitly
but rather quote only the form of the heavy baryon limit. Due to its length
the entire expression of the heavy fermion limit is relegated
to app.~\ref{app.b}. In the present 
section we will refer only  to the counterterms
which are needed for our calculation. The reason for the choice of these 
terms will become clear below when we estimate the LECs by means of 
the resonance saturation principle.

The pseudoscalar Goldstone fields ($\phi = \pi, K, \eta$) are collected in
the  $3 \times 3$ unimodular, unitary matrix $U(x)$, 
\begin{equation}
 U(\phi) = u^2 (\phi) = \exp \lbrace 2 i \phi / \Fnod \rbrace
\end{equation}
with $\Fnod$ being the pseudoscalar decay constant (in the chiral limit), and
\begin{eqnarray}
 \phi =  \frac{1}{\sqrt{2}}  \left(
\matrix { {1\over \sqrt 2} \pi^0 + {1 \over \sqrt 6} \eta
&\pi^+ &K^+ \nonumber \\
\pi^-
        & -{1\over \sqrt 2} \pi^0 + {1 \over \sqrt 6} \eta & K^0
        \nonumber \\
K^-
        &  \bar{K^0}&- {2 \over \sqrt 6} \eta  \nonumber \\} 
\!\!\!\!\!\!\!\!\!\!\!\!\!\!\! \right) \, \, \, \, \, . 
\end{eqnarray}
Under SU(3)$_L \times$SU(3)$_R$, $U(x)$ transforms as $U \to U' =
LUR^\dagger$, with $L,R \in$ SU(3)$_{L,R}$.
The matrix $B$ denotes the baryon octet, 
\begin{eqnarray}
B  =  \left(
\matrix  { {1\over \sqrt 2} \Sigma^0 + {1 \over \sqrt 6} \Lambda
&\Sigma^+ &  p \nonumber \\
\Sigma^-
    & -{1\over \sqrt 2} \Sigma^0 + {1 \over \sqrt 6} \Lambda & n
    \nonumber \\
\Xi^-
        &       \Xi^0 &- {2 \over \sqrt 6} \Lambda \nonumber \\} 
\!\!\!\!\!\!\!\!\!\!\!\!\!\!\!\!\! \right)  \, \, \, ,
\end{eqnarray}
which under $SU(3)_L \times SU(3)_R$ transforms as any matter field,
\begin{equation} 
B \to B' = K \, B \,  K^\dagger
 \, \, \, ,
\end{equation}
with $K(U,L,R)$ the compensator field representing an element of the
conserved subgroup SU(3)$_V$.
To the order we are working the effective Lagrangian has the form
\beq
{\cal L}_{\mbox{eff}}  =  \: {\cal L}_{\phi B} \: + \: 
        {\cal L}_{\phi B}^W  \:    + \: {\cal L}_{\phi}
        \:    + \: {\cal L}_{\phi}^W \qq ,
\eeq
where $ {\cal L}_{\phi} = {\cal L}_{\phi}^{(2)} + {\cal L}_{\phi}^{(4)}$ 
is the usual (strong and electromagnetic) mesonic 
Lagrangian up to fourth chiral order , see {\it e.g.} \cite{GL1}. 
\footnote{The fourth order is needed for the $Z$-factor of the pion.}
From the weak mesonic Lagrangian only the term
\beq \label{weak}
{\cal L}_{\phi}^W = \frac{\Fnod^2}{4} \, h_{\pi} \, {\rm tr}
      \Big( h_+ u_{\mu} u^{\mu}\Big)
\eeq
contributes to the order we are working. 
Here, we have defined
\beq
h_+ = u^{\dagger} h u +  u^{\dagger} h^{\dagger} u  \qquad , 
\eeq
with
$h^{a}_{b} = \delta^{a}_{2} \delta^{3}_{b}$ being
the weak transition matrix. 
Note that $h_+$   transforms as a matter field.
The weak coupling $h_{\pi}$ is well-determined from
weak kaon decays -- $h_{\pi} = 3.2 \times 10^{-7}$.

For the strong meson-baryon Lagrangian ${\cal L}_{\phi B}  $
one writes
\beqa
{\cal L}_{\phi B} &=&
 {\cal L}_{\phi B}^{(1)} + {\cal L}_{\phi B}^{(2)} \qq ,    
\eeqa
where the superscript denotes the chiral order and 
\beqa
{\cal L}_{\phi B}^{(1)}  
& = &
 \ci \, \tr \Big( \bar{B} [ v \cdot D , B] \Big) +            
D \, \tr \Big( \bar{B} S_{\mu} \{ u^{\mu}, B\} \Big) 
+ F \, \tr \Big( \bar{B} S_{\mu} [ u^{\mu}, B] \Big) 
\eeqa
\beqa
{\cal L}_{\phi B}^{(2)}  & = & {\cal L}_{\phi B}^{(2,rc)} \no \\
& = &
- \frac{1}{2 \mnod} \tr \Big( \bar{B} [ D_{\mu}, [D^{\mu}, B]] \Big) 
+ \frac{1}{2 \mnod} \tr \Big( \bar{B} [ v \cdot D, [v \cdot D, B]] \Big) 
\eeqa
with $2 S_\mu = \ci \gamma_5 \sigma_{\mu \nu} v^{\nu}$ 
denoting the 
Pauli--Lubanski spin vector.
For $ {\cal L}_{\phi B}^{(3)}  $ we only consider the part which 
renormalizes the $Z$-factors.
\beqa
{\cal L}_{\phi B}^{(3)} & = &
\ci x_1 \tr \Big( \bar{B} \{ \chi_+ , [ v \cdot D , B ]\} \Big) 
+
\ci x_2 \tr \Big( \bar{B} [ \chi_+ , [ v \cdot D , B ]] \Big) \no \\
& + &
\ci x_3 \tr \Big( \bar{B} [ v \cdot D , B ] \Big) \tr \Big(\chi_+\Big)   
\eeqa

We do not include the part of
$ {\cal L}_{\phi B}^{(3)}  $ which renormalizes the axial-vector
couplings since already a simple lowest order fit for those 
yielding $ D \simeq 3/4 $ and $ F \simeq 1/4 $,
which are the values in the $SU(6)$ limit, 
gives a very satisfactory description.
There do not appear additional unknown LECs.

Having dealt with its strong counterpart,
the weak meson-baryon Lagrangian ${\cal L}_{\phi B}^W  $ reads
\beq
{\cal L}_{\phi B}^W  \:  = \:
{\cal L}_{\phi B}^{W \, (0)}  \, +\,
{\cal L}_{\phi B}^{W \, (1)}  \, +\,
{\cal L}_{\phi B}^{W \, (2)}   \qq . 
\eeq
The form of the lowest order 
Lagrangian is
\beq
{\cal L}_{\phi B}^{W \, (0)}  =  \:
d \, \tr \Big( \bar{B}  \{ h_+ , B\} \Big) + \:
f \, \tr \Big( \bar{B}  [ h_+ , B ] \Big) \qquad ,
\eeq
and these are the {\it only} terms considered in previous 
calculations, \cite{DGH,DoGoHo,Bij,Jen}.
To next order there is no contribution if we resort to resonance exchange only.
In order to achieve a much better fit to the decay amplitudes, however,
we have to
include further counterterms, which by standard arguments
should be numerically more significant the
lower the chiral order. 
Thus we take the counterterms
in ${\cal L}_{\phi B}^{W \, (1)}$ into account, see app.~B. This is also 
indicated in \cite{Neu} where 
a rough estimate of the LECs of the weak baryon Lagrangian
of order ${\cal O}(p)$ has been given using the weak deformation model and 
the author comes to the conclusion that one cannot understand
nonleptonic hyperon decays without such LECs.
There are two types of terms. The terms involving the LECs $g_3$ to $g_{10}$
({\it cf}. app.~B) have the structure $\bar{B} h_+ \, v \cdot u \, B$. 
Their contributions are proporional to $v \cdot k$ with $k$ being the meson
four--momentum. The term $v \cdot k$ is the energy of the meson in the rest 
frame of the heavy baryon, {\it i.e.} $v_{\mu} = (1,0,0,0)$,
and can be expressed as the difference of the squared masses of the 
external baryons. Since such mass differences are to lowest order 
analytic in the quark masses they can be absorbed by explicit
symmetry breaking terms in ${\cal L}_{\phi B}^{W \, (2)}$ of the 
form $\bar{B} h_+ \chi_+ B$. In the following then we work only with 
the remaining counterterms and will omit the terms
$g_3$ to $g_{10}$.
This leaves us with the following Lagrangian 
${\cal L}_{\phi B}^{W \, (1)}$ at first order
\beqa
{\cal L}_{\phi B}^{W \, (1)}
& = & 
2 g_{11} \bigg\{ \tr \Big( \bar{B} S_{\mu}  
                             [ h_+ , [u^{\mu}, B] ] \Big) +
     \tr \Big( \bar{B} S_{\mu}  
                         [ u^{\mu} , [h_+, B] ] \Big) \bigg\} \no \\
& + & 
2 g_{13} \bigg\{ \tr \Big( \bar{B} S_{\mu}  
                            [ h_+ , \{ u^{\mu}, B\} ] \Big) +
  \tr \Big( \bar{B} S_{\mu}  
                      \{ u^{\mu} , [h_+, B] \} \Big) \bigg\} \no \\
& + & 
2 g_{15} \bigg\{ \tr \Big( \bar{B} S_{\mu}  
                               \{ h_+ , [u^{\mu}, B] \} \Big) +
  \tr \Big( \bar{B} S_{\mu}  
                          [ u^{\mu} , \{ h_+, B\} ] \Big) \bigg\} \no \\
& + & 
2 g_{16} \bigg\{ \tr \Big( \bar{B} h_+ \Big) S_{\mu}  
                              \tr \Big( u^{\mu} B \Big)
+ \tr \Big(\bar{B} u^{\mu}\Big) S_{\mu}   
                                 \tr \Big( h_+ B \Big) \bigg\} \no \\ 
& + & 
2 g_{18} \tr \Big( \bar{B} S_{\mu}  B \Big) 
                             \tr \Big( u^{\mu} h_+ \Big)
\eeqa

However, in second order there appear explicit symmetry 
breaking terms besides the double--derivative terms  
\beq
{\cal L}_{\phi B}^{W \, (2)}   
=  
{\cal L}_{\phi B}^{W \, (2,br)}  \: +  \:
\sum_{i} \, h_i O_i^{(2)} 
\eeq
with
\beqa
{\cal L}_{\phi B}^{W \, (2,br)}  
& = &
h_3 \bigg\{ \tr \Big( \bar{B}  [ h_+ , [ \chi_+, B] ] \Big)  
         +\tr \Big( \bar{B}  [ \chi_+, [ h_+ , B ]]  \Big) \bigg\} \no \\ 
& + & 
h_5 \bigg\{ \tr \Big( \bar{B}  [ h_+ , \{ \chi_+, B\} ] \Big)  
         +\tr \Big( \bar{B}  \{ \chi_+, [ h_+ , B ]\}  \Big) \bigg\} \no \\ 
& + & 
h_7 \bigg\{ \tr \Big( \bar{B}  \{ h_+ , [ \chi_+, B] \} \Big)  
         +\tr \Big( \bar{B}  [ \chi_+, \{ h_+ , B \} ]  \Big) \bigg\} \no \\ 
& + & 
h_8 \bigg\{ \tr \Big( \bar{B} h_+ \Big)  \tr \Big( \chi_+  B \Big)
+ \tr \Big(\bar{B} \chi_+ \Big) \tr \Big( h_+ B \Big)
\bigg\} \no \\ 
& + & 
h_{11} \tr \Big( \bar{B}  [ h_+ , B ] \Big) \tr \Big( \chi_+ \Big) +
h_{12} \tr \Big( \bar{B}  \{ h_+ , B ]\} \Big) \tr \Big( \chi_+ \Big) 
\eeqa
Here $\chi_+ = u^\dagger
\chi u^\dagger + u \chi^\dagger u$ is proportional to the quark mass
matrix ${\cal M}$  = ${\rm diag}(m_u,m_d,m_s)$, since $\chi = 2 B {\cal
  M}$. Also, $B = - \langle 0 | \bar{q} q | 0 \rangle / \Fnod^2$ is
the order parameter of the spontaneous symmetry violation,
and we assume $B \gg \Fnod$. 
From the entire list of the chiral order two double--derivative terms
it turns out that only
two terms need to be retained
\beq \label{mon2}
\sum_{i} \, h_i O_i^{(2)} = 
h_1 \, \tr \Big( \bar{B}  [ [ D_{\mu}, [D^{\mu}, h_+]] , B ] \Big) +
h_2 \, \tr \Big( \bar{B}  \{ [ D_{\mu}, [D^{\mu}, h_+]] , B \} \Big) 
\eeq
The relativistic corrections do not contribute in this order. 
Since we choose the four velocity $ v_{\mu}
= (1,0,0,0)$, {\it i.e.} the rest frame of the decaying baryon,
the derivative on the incoming baryon field
carries  the velocity $ v$. Therefore, some of the terms vanish, because
$ S \cdot v = 0 $ or since they are proportional to
$ \epsilon_{\mu \nu \alpha \beta} \, v^{\alpha} \,
v^{\beta}$.
On the other hand, the energies of the external particles 
in the heavy baryon formalism can be expressed as the difference 
of the squared masses of the external baryons and 
since such mass differences are to lowest order analytic in the quark masses
they count as chiral order two. Thus terms with two derivatives $ v \cdot D$
can be neglected. 

We can further reduce the number of independent counterterms, since
$ h_{11} $ and  $ h_{12} $ of ${\cal L}_{\phi B}^{W \, (2,br)} $
amount to quark mass renormalizations of $d$ and $f$ in
${\cal L}_{\phi B}^{W \, (0)}$. To be specific, one can 
absorb the effects of
$ h_{11} $ and  $ h_{12} $ in $d$ and $f$ as follows
\beq
d  \rightarrow d - h_{12}^r \, \tr (\chi_+) \qq , \qq
f  \rightarrow f - h_{11}^r \, \tr (\chi_+) \qq ,
\eeq  
where the superscript $ r $ denotes the finite remainder of the LECs
after renormalization, since the infinite pieces of $ h_{11}$ and $ h_{12} $ 
cancel the divergences arising from the loop diagrams. After 
that one absorbs the finite remainder in the
phenomenological values of $d$ and $f$.
This is a very general feature of CHPT calculations in higher
orders. For example, in $\pi \pi$ scattering there exist six LECs at two
loop order $(q^6)$ \cite{BCEGS}, but only two new independent terms
$\sim s^3$ and $\sim s \, M_\pi^4$. The other four LECs simply make the
${\cal O}(q^4)$ counter terms $\bar{\ell}_i$ ($i=1,2,3,4)$ quark 
mass-dependent. 

Here we lump the lower and higher order terms together in order to minimize
the number of independent couplings.
Consider furthermore the terms  $ h_{1} $ and  $ h_{2} $ in eq.~(\ref{mon2}).
To the order  we are working one can therein
replace the covariant derivatives by 
the partial ones. Then because 
$ k^2 = M_{\pi}^2 $, with $ k $
the momentum  of the outgoing pion, $ h_{1} $ and  $ h_{2} $ can also be 
absorbed into $d$ and $f$.
So we end up with the familiar
two unknown counterterms in lowest order and just
eight in the next two orders.

\subsection{Estimation of the low energy constants}
Performing the calculations with the complete Lagrangian of app.~\ref{app.b},
one has, of course, no predictive power. 
Indeed there exist only eight experimental results:
the s- and p-wave amplitudes for the four independent hyperon decays.
On the other side, the theoretical predictions contain considerably more 
than eight low 
energy constants.
Clearly, we are not able to fix all the low--energy
constants appearing in ${\cal L}_{\phi B}^W$
from data, even if we resort to large $N_c$ arguments.
We will therefore use the principle of resonance saturation in order to
estimate the importance of
these constants, which works very accurately in the meson
sector \cite{reso,reso1,reso2} and also in the
baryon sector \cite{BM}. In the baryon case, one has to account for 
excitations of meson ($R$) and baryon ($N^*$) resonances. One writes
down the effective Lagrangian with these resonances chirally coupled to
the Goldstone bosons and the baryon octet, calculates the  Feynman 
diagrams pertinent to the process under consideration and, finally,
lets the resonance masses become infinite (with fixed ratios of
coupling constants to masses). This generates higher order terms in
the effective meson--baryon Lagrangian with coefficients expressed in
terms of a few known resonance parameters. Symbolically, we can write
\begin{equation}
\tilde{{\cal L}}_{\rm eff} [\, U,B,R,N^* \, ] 
\stackrel{m_R , m_{N^*} \to \infty }{\longrightarrow}
{\cal L}_{\rm eff} [\, U,B \, ]  \qq . 
\end{equation}
It is important to stress that {\it only} after integrating out the heavy
degrees of freedom from the effective field theory is one allowed to
perform the heavy mass limit for the ground--state baryon octet.
Assuming that the spin-3/2 decuplet states are the main contributions 
to the LECs, which is, {\it e.g.}, the case in the self-energy diagrams
of the baryon octet, \cite{BM}, we will treat our results  
as being only indicative.
This is for several reasons.
On the one hand, there exist many higher baryon resonances, {\it e.g.},
the parity--even spin--1/2 octet which includes the Roper $N^* (1440)$.
Also, it is important to
stress that for the resonance contribution to the baryon masses, one
has also to include Goldstone boson loops, 
since there are no tree level diagrams contributing
to the processes under consideration. This is different from the
situation as in form factors or scattering processes.
Treating these resonances relativistically leads to some complications 
that have already been discussed in \cite{BM}:
\begin{enumerate}

\item[$\circ$] First, terms arise which are non--analytic in the
meson masses. Clearly, to avoid any double counting and 
to be consistent with the
requirements of analyticity, one should only consider the analytic
terms in the meson masses generated by such loop diagrams. 
Here, we only have to consider the terms up-to-and-including
second chiral order which are linear in the quark masses or 
quadratic in the external momenta. Since the lowest nonanalytic contributions
in these resonance diagrams appear at fourth order, we do not have to bother 
about this.

\item[$\circ$] Second, to the order we are working, 
the analytic
pieces are divergent .
Therefore, we can only determine the analytic
resonance contribution up to renormalization constants.
clearly then we do not obtain an explicit numerical result for the LECs.

\item[$\circ$] Third, since the baryon excitations are treated
  relativistically, as explained
above, there does not exist strict power counting \cite{GSS} and thus 
one must include higher loop diagrams.

\end{enumerate}
All the arguments
mentioned above suggest that this scheme can be used only to 
decide which LECs derive important contributions from resonances.
We will thus use the results from the resonance diagrams in the following 
manner:
If such diagrams do not contribute to a specific LEC we neglect
this constant. The remaining LECs will be kept in our calculations 
as unknown parameters to be fixed from experiment.

Consider now the decuplet contribution. We treat these fields
relativistically and only at the last stage let the mass become very
large. The interaction Lagrangian between the spin--3/2
fields (denoted by $\Delta$), the baryon octet and the Goldstone bosons reads
\begin{equation}
{\cal L}_{\Delta B \phi} = 
\frac{{\cal C}}{2} \, \biggl\{ \bar{\Delta}^{\mu ,abc} \, 
 \Theta_{\mu \nu} (Z) \, (u^\nu)_a^i \, B_b^j \, 
\epsilon_{cij}- \bar{B}^b_i \, (u^\nu)^a_j \,  \Theta_{\nu \mu} (Z) \,
{\Delta}^\mu_{abc} \,\epsilon^{cij}\, \biggr\} \, \,  \, ,
\label{lmbd}
\end{equation}
where $a, b, \ldots , j$ are SU(3)$_f$ indices
and the coupling constant $1.2 < {\cal C} < 1.8 $ can be
determined from the decays $\Delta \to
B \pi$. The Dirac matrix operator $\Theta_{\mu \nu} (Z)$ is given by
\begin{equation}
\Theta_{\mu \nu} (Z) = g_{\mu \nu} - \biggl(Z + \frac{1}{2} \biggr) \,
\gamma_\mu \, \gamma_\nu \, \, \, \, .
\label{theta}
\end{equation}
For the off--shell parameter $Z$, we use $Z = -0.3$ from the
determination of the $\Delta$ contribution to the $\pi N$ scattering
volume $a_{33}$ \cite{armin}. (This value is also consistent with recent
studies of $\Delta (1232)$ contributions to the nucleon
electromagnetic polarizabilities \cite{bkmzas} and to threshold 
pion photo-- and electroproduction \cite{bkmpe}. )
For the  processes to be discussed,
we require only the lowest order form of $u_\mu$,
\begin{equation}
(u_\mu)^i_a = -\frac{2}{F_\pi} \partial_\mu \, \phi_a^i + 
{\cal O}(\phi^2) \quad .
\end{equation}
The propagator of the spin--3/2 fields is
\begin{equation}
G_{\beta \delta} (p) = -i\frac{p \!\!/ + m_\Delta}{p^2-m_\Delta^2} 
\, \biggl( g_{\beta \delta} - \frac{1}{3} \gamma_\beta \gamma_\delta -
\frac{2 p_\beta p_\delta}{3m_\Delta^2} + \frac{p_\beta \gamma_\delta -
p_\delta \gamma_\beta}{3 m_\Delta} \, \biggr) \, \, \, ,
\end{equation}
with $m_\Delta = 1.38$~GeV being the average decuplet mass.
Furthermore, we need the weak strangeness changing Lagrangian for the
decuplet fields
\beq
{\cal L}_{\Delta \phi}^W  =
\, h_c \, \bar{\Delta}^{\mu, abc} (h_+)_a^i \Delta_{\mu, ibc}
\eeq
We can now evaluate the diagram  shown in fig.~1. 
With the labeling of
the momenta as in the figure, this leads to
\begin{equation}
I_\Delta (p,q) = \frac{- \,{\cal C}^2 \, h_c }{\sqrt{2}\,
2F^2_\pi} \int \frac{d^4
k}{(2\pi)^4} \frac{l^\sigma \,
\Theta_{\sigma \rho} (Z) \, G^{\rho \mu} (q+l) \, G_{\mu \nu} (p+l)
\, \Theta^{\nu \lambda} (Z) l_\lambda }{l^2-M_a^2+i\epsilon} \qq ,
\label{Idelta}
\end{equation}
where the relevant Clebsch--Gordan coefficient has been omitted and $M_a$
is the mass of the meson in the loop. 
This integral is evaluated on the mass--shell of the
external baryons, i.e at $p \!\!/ = q \!\!/ = \mnod$, and
splits into various contributions according to the power of momenta in
the numerator and the number of propagators. Each such term is then
expanded in powers of Goldstone boson masses up-to-and-including
${\cal O}(M_a^2)$ for $I_\Delta (p,q) $.
Only then is the large mass limit of the decuplet 
taken. This then gives the contribution to the various LECs.
Assuming analyticity of the integral with respect to the external
momenta $ p $ and $ q$ one can expand in terms of 
the momentum transfer squared $ t= ( p - q)^2$. This would amount
to the following expansion in the quark masses and $t$ for the integral
\beq
I_\Delta (p,q)  =   a + b M_a^2 + c t + \ldots \qq ,
\eeq
where the ellipsis stand for higher orders and 
$a,b,c$ are constants. 
The term $ct$ in the expansion for $I_\Delta (p,q) $  is included
in the terms with the LECs $h_1$ and $h_2$ in eq.(\ref{mon2}).
Since these effects are absorbed by $d$ and $f$ as explained above, 
it is sufficient to
evaluate the integral for $t=0$, i.e. $p=q$.
We can now work out the complete integral at $p \! \! / =
q \! \! / = \mnod$ and find
\beqa  \label{Ideltaresz}
I_\Delta  & = &
\ci \frac{2}{9} \, (Z + 1)\, \bigg\{ \,
7 \, \Big[ \, 2\, L + \frac{1}{16 \, \pi^2} \ln \left( 
\frac{m_{\Delta}^2}{\lambda^2} \right) \Big] + \frac{3}{8 \, \pi^2} 
\bigg\} \, m_{\Delta} \, \mnod  + \ldots \no \\
&  &
+ \ci \frac{2}{9} \, (Z + 1)\, \bigg\{ \,
2\, L + \frac{1}{16 \, \pi^2} \ln \left( 
\frac{m_{\Delta}^2}{\lambda^2} \right)+ \frac{3}{8 \, \pi^2} 
\bigg\} \, \frac{\mnod}{m_{\Delta}} M^2_a + \ldots \eeqa
where the ellipsis stand for subleading orders in the
$1/ m_{\Delta}$ expansion.
One notices  that in this relativistic
treatment, the dimension zero and two  LECs
are not  finite (the dimension zero LECs {\it are} finite
in the heavy baryon approach) \cite{GSS}. 
The structure of eq.(\ref{Ideltaresz}) indicates that the first and second
term of $ I_\Delta $
contributes to $ d, f $ and the $h_{3,5,7,8,11,12}$, respectively.
So we end up
with the Lagrangian given in the previous section except those terms in
${\cal L}_{\phi B}^{W \, (1)}$.

\section{Non-leptonic hyperon decays}
Having constructed the relevant building blocks, we can now get down to work.
The matrix elements for non-leptonic hyperon decay are written as
\beq
{\cal A}( B_i \rightarrow B_j \, \pi) =
\bar{u}_{B_j} \Big\{ \, A_{ij}^{(S)} + \, A_{ij}^{(P)}\gamma_5 \Big\}u_{B_i}
\eeq
where $ A_{ij}^{(S)}$ is the parity-violating s-wave amplitude and
$ A_{ij}^{(P)}$ is  the corresponding parity-conserving p-wave term.
In the heavy baryon formulation the p-wave must be modified, since 
$\gamma_5$ connects the light with the heavy degrees of freedom which are 
integrated out in this scheme.
One therefore introduces the modified heavy baryon p-wave amplitude
${\cal A}_{ij}^{(P)} $ by
\beq
A_{ij}^{(P)} = - \frac{1}{2} (E_j + m_j) {\cal A}_{ij}^{(P)} \qq ,
\eeq
where $E_j$ and $ m_j$ are the energy and mass of the outgoing baryon, 
respectively.
In the rest frame of the heavy baryon, $ v_{\mu} = (1,0,0,0) $,
the decay amplitude reduces to the non-relativistic form
\beqa
{\cal A}( B_i \rightarrow B_j \, \pi) & = &
\bar{\chi}_{B_j} \, \Big\{ \, {\cal A}_{ij}^{(S)} + \, 
\frac{1}{2}\,  \vec{k} \cdot \vec{\sigma}\,  {\cal A}_{ij}^{(P)}\, 
\Big\}\chi_{B_i}\no \\
& = &
\bar{\chi}_{B_j} \, \Big\{ \, {\cal A}_{ij}^{(S)} + \, 
S \cdot k \, {\cal A}_{ij}^{(P)}\,  \Big\}\chi_{B_i} \qq ,
\eeqa
where $k$ is the outgoing momentum of the pion and $ S_\mu $
is the Pauli-Lubanski spin vector, which
in the rest frame is given by
$S^{\mu}_{\vec{v}=0} =  ( 0 , \frac{1}{2} \vec{\sigma} ) $.
Isospin symmetry of the strong interactions implies the relations 
\beqa \label{iso}
&&
{\cal A}(\Lambda \rightarrow p \, \pi^-)
+ \sqrt{2} \, {\cal A}(\Lambda \rightarrow n \, \pi^0) = 0 \no \\
&&
{\cal A}(\Xi^- \rightarrow  \Lambda \, \pi^-)
+ \sqrt{2} \, {\cal A}(\Xi^0 \rightarrow  \Lambda \, \pi^0) = 0 \no \\
&&
\sqrt{2} \, {\cal A}(\Sigma^+ \rightarrow p \, \pi^0)
+ {\cal A}(\Sigma^- \rightarrow n \, \pi^-)
- {\cal A}(\Sigma^+ \rightarrow n \, \pi^+) = 0
\eeqa
which hold both for s- and p-waves.
We choose
$\Sigma^+  \rightarrow n \, \pi^+ \, , \, 
\Sigma^- \rightarrow n \, \pi^- \, , \, 
\Lambda \rightarrow p \, \pi^- \, \mbox{and} \,
\Xi^- \rightarrow  \Lambda \, \pi^- $
to be the four independent decay amplitudes which are not related by isospin.

We calculate all tree and one loop diagrams contributing to these
processes by making use of the Lagrangian from the previous section.
For the p-waves we have to consider pole diagrams, which leads to
some difficulties with the usual chiral counting scheme. Thus consider 
the inverse
of the free
propagator of the internal baryon, which is either $ v \cdot p$ or 
$ v \cdot q$ with $p$ and $q$ being the off-shell momenta of the incoming
or outcoming baryon, respectively. 
For example, in the rest frame of the decaying baryon the kinetic energy
of the outgoing baryon may be written as
\beq \label{vq}
v \, \cdot \, q = \frac{1}{2 \, m_i} 
             \Big( m_i^2 + m_j^2 -2 \mnod m_i - M_{\pi}^2 \Big) \qq .
\eeq 
Since the baryon masses are analytic to linear order in the quark masses
we see that $v \, \cdot \, q  = {\cal O}(p^2)$, as noted
in the previous section.

The general structure of the s-wave decay amplitudes is
\beqa   \label{swa}
{\cal A}_{ij}^{(s)} & = &
\frac{1}{\sqrt{2}\, F_{\pi}}\, \bigg\{ \,
\alpha_{ij}^{(s)} \, + \,
\beta_{ij}^{(s)\,Q} \, M_Q^2 \, + \, 
\frac{1}{\Lambda_\chi^2} \, \gamma_{ij}^{(s)\,Q} M_Q^2 \, \ln \left(
\frac{ M_Q^2}{\mu^2} \right) 
+ \frac{1}{\Lambda_\chi^2} \, \alpha_{ij}^{(s)} \, \lambda_{ij} \bigg\}
\eeqa
where $ Q = \pi, K, \eta$ and $\mu$ represents the 
dimensional regularization scale.
Also, $\Lambda_\chi = 4 \pi F_\pi$ represents the scale
of chiral symmetry breaking and arises naturally during the evaluation
of the loop integrals.
The coefficient $\alpha_{ij}$ is the tree level result, while
$\beta_{ij}$ contains both the contributions from the second order 
counterterms and the analytic parts of the loop diagrams, and
$\gamma_{ij}$ summarizes the non-analytic loop pieces.
Finally, we have the modifications arising from the
multiplication of the tree result with 
the wavefunction renormalization $Z$-factors and from 
replacing the pseudoscalar decay constant in the chiral
limit $ \Fnod$ by the physical pion decay constant $F_\pi$.
Since we do not include  self-energy corrections 
of the external particles explicitly, we have to multiply the decay
amplitudes by the relevant $Z$-factors. 
To be specific, the quantity $ \lambda_{ij}$ is defined as follows
\beq
\frac{1}{\Lambda_\chi^2} \, \lambda_{ij} =
\frac{1}{2} (Z_i -1) + \frac{1}{2} (Z_j -1) + \frac{1}{2} (Z_\pi -1) 
+ \delta F_\pi  \qq ,
\eeq
where the specific expressions for the $Z$-factors and $\delta F_\pi $ can 
be found in app.~\ref{app.c}. 

The diagrams that contribute to the s-waves are shown in fig.~2.
Here the loop diagrams are divergent and have to be renormalized 
by appropriate counterterms. The renormalization procedure is
outlined in the next section. The decay amplitudes are then expressed
in terms of the finite remainder of the LECs. For notational
simplicity, we will use 
the same symbol for the finite remainder of these LECs by neglecting
the superscript $r$. That is, {\it e.g.}, $h_3$
is actually $h_3^r$, where $h_3^r$ is
the finite remainder defined in the next section.
Furthermore, we count the external momenta multiplied by the 
four--velocity $v$ as effectively of order
${\cal O}(p^2)$, which allows us to write the analytic results of the loop 
integrals in a more compact form by neglecting higher order parts.
The results then read
\beqa
\alpha_{\Sigma^- n}^{(s)} & = &  d -f 
\no \\
\beta_{\Sigma^- n}^{(s)\,\pi} & = & -4 h_3 -4 h_5 + 4 h_7 
+ \frac{1}{\Lambda_\chi^2} \, (D+F) \, \Big(2F(d-f) - \frac{1}{3}D(d+3f)\Big)
\no \\
\beta_{\Sigma^- n}^{(s)\,K} & = & 4 h_3 -4 h_5 - 4 h_7 +
  \frac{1}{\Lambda_\chi^2} \, (D+F) \, 
            \Big( - \frac{1}{2} (D-F)(d+f) + \frac{1}{6}(D+3F)(d-3f)\Big)
\no \\
\beta_{\Sigma^- n}^{(s)\,\eta} & = & - \frac{1}{\Lambda_\chi^2} \,
                            \frac{1}{3}(D-3F)D(d-f)
\no \\
\gamma_{\Sigma^- n}^{(s)\,\pi} & = &  \frac{7}{24} (d-f)
  + \frac{3}{2} \, (D+F) \, \Big(2F(d-f) - \frac{1}{3}D(d+3f)\Big)
\no \\
\gamma_{\Sigma^- n}^{(s)\,K} & = &  - \frac{5}{12} (d-f) +
\frac{3}{2} \,(D+F) \, 
            \Big( - \frac{1}{2} (D-F)(d+f) + \frac{1}{6}(D+3F)(d-3f)\Big)
\no \\
\gamma_{\Sigma^- n}^{(s)\,\eta} & = & - \frac{3}{8} (d-f) 
                            - \frac{1}{2}(D-3F)D(d-f)\no \\
&& \no \\
\alpha_{\Lambda p }^{(s)} & = &  - \frac{1}{\sqrt{6}} (d+3f)\no \\
\beta_{\Lambda p }^{(s)\,\pi} & = &  4 \sqrt{\frac{3}{2}} 
     \Big( - h_3 + \frac{1}{3} h_5 - \frac{1}{3} h_7 + \frac{2}{3} h_8 \Big) 
      +\frac{1}{\Lambda_\chi^2} \,\frac{3}{\sqrt{6}}(D+F)D (d-f) \no \\
\beta_{\Lambda p }^{(s)\,K} & = &  - 4 \sqrt{\frac{3}{2}} 
     \Big( - h_3 + \frac{7}{3} h_5 - \frac{1}{3} h_7 + \frac{2}{3} h_8 \Big)
\no \\
& - &
\frac{1}{\Lambda_\chi^2} \,\frac{1}{\sqrt{6}}
\Big( \frac{3}{2} (D-F) (D-3F)(d+f) + \frac{1}{6}(D+3F)(D-3F)(d-3f)
\Big)\no \\
\beta_{\Lambda p }^{(s)\,\eta} & = &  - \frac{1}{\Lambda_\chi^2} \,
\frac{1}{3\sqrt{6}}(D-3F) D (d+3f)\no \\
\gamma_{\Lambda p }^{(s)\,\pi} & = & 
    - \frac{1}{\sqrt{6}} (d+3f)\frac{7}{24} 
     + \frac{9}{2\sqrt{6}} \, (D+F)D (d-f)\no \\
\gamma_{\Lambda p }^{(s)\,K} & = & 
           \frac{1}{\sqrt{6}} \frac{5}{12} (d+3f) \no \\
& - &
\frac{3}{2\sqrt{6}}
\Big( \frac{3}{2} (D-F) (D-3F)(d+f) + \frac{1}{6}(D+3F)(D-3F)(d-3f)
\Big)\no \\
\gamma_{\Lambda p }^{(s)\,\eta} & = & 
              + \frac{1}{\sqrt{6}} \frac{3}{8} (d+3f)
              - \frac{1}{2\sqrt{6}} (D-3F) D (d+3f)\no \\
&& \no \\
\alpha_{\Xi^- \Lambda }^{(s)} & = &  - \frac{1}{\sqrt{6}} (d-3f)\no \\
\beta_{\Xi^- \Lambda }^{(s)\,\pi} & = &  4 \sqrt{\frac{3}{2}} 
     \Big( - h_3 - \frac{1}{3}  h_5 + \frac{1}{3} h_7 + \frac{2}{3} h_8 \Big) 
      +\frac{1}{\Lambda_\chi^2} \,\frac{3}{\sqrt{6}}(D-F)D (d+f) \no \\
\beta_{\Xi^- \Lambda }^{(s)\,K} & = &  - 4 \sqrt{\frac{3}{2}} 
     \Big( - h_3 - \frac{7}{3} h_5 + \frac{1}{3} h_7 + \frac{2}{3} h_8 \Big)
\no \\
& - &
\frac{1}{\Lambda_\chi^2} \,\frac{1}{\sqrt{6}}
\Big( \frac{3}{2} (D+F) (D+3F)(d-f) + \frac{1}{6}(D+3F)(D-3F)(d+3f)
\Big)\no \\
\beta_{\Xi^- \Lambda }^{(s)\,\eta} & = &  - \frac{1}{\Lambda_\chi^2} \,
\frac{1}{3\sqrt{6}}(D+3F) D (d-3f)\no \\
\gamma_{\Xi^- \Lambda }^{(s)\,\pi} & = & 
    - \frac{1}{\sqrt{6}} (d-3f)\frac{7}{24} 
     + \frac{9}{2\sqrt{6}} \, (D-F)D (d+f)\no \\
\gamma_{\Xi^- \Lambda }^{(s)\,K} & = & 
           \frac{1}{\sqrt{6}} \frac{5}{12} (d-3f) \no \\
& - &
\frac{3}{2\sqrt{6}}
\Big( \frac{3}{2} (D+F) (D+3F)(d-f) + \frac{1}{6}(D+3F)(D-3F)(d+3f)
\Big)\no \\
\gamma_{\Xi^- \Lambda}^{(s)\,\eta} & = & 
              + \frac{1}{\sqrt{6}} \frac{3}{8} (d-3f)
              - \frac{1}{2\sqrt{6}} (D+3F) D (d-3f)\no \\
\eeqa
All the other coefficients in eq.(\ref{swa}) vanish.

For the p-waves one has the form
\beqa    \label{pwa}
{\cal A}_{ij}^{(p)} & = &
\frac{1}{\sqrt{2}\, F_{\pi}}\, \bigg\{ \,
\alpha_{ij}^{(p)} \, + \,
\beta_{ij}^{(p)\,Q} \, M_Q^2 \, + \, 
\frac{1}{\Lambda_\chi^2} \, \gamma_{ij}^{(p)\,Q} M_Q^2 \, \ln \left(
\frac{ M_Q^2}{\lambda^2} \right) \no \\
&&
+ \epsilon_{ij}^{(p)}
+ \frac{1}{\mnod} v \cdot k \, \delta_{ij}^{(p)}
+ \frac{1}{\mnod} \rho_{ij}^{(p)}
+ \frac{1}{\Lambda_\chi^2} \, \alpha_{ij}^{(p)} \, \lambda_{ij} 
+ \frac{1}{2} \, h_{\pi} \, \frac{M_{\pi}^2}{M_{\pi}^2-M_K^2}\, \phi_{ij}^{(p)}
\bigg\}
\eeqa
The $\epsilon_{ij}^{(p)}$ are the contributions of the 
counterterms $g_{11}$ to $g_{16}$ of the weak Lagrangian
${\cal L}_{\phi B}^{W \, (1)}$. 
Both the terms $\delta_{ij}^{(p)}$ and $\rho_{ij}^{(p)}$ 
arise from additional $1/\mnod$ corrections 
appearing for the
p-waves as described in app.~\ref{app.b} whereas $\phi_{ij}^{(p)}$
is the contribution from the weak decay of the meson.

The diagrams which contribute to p-waves are depicted in fig.~3 and fig.~4.
(Note that diagrams ~3{\it l},  ~3$\,$m have not been considered
in previous calculations.)
In most of the diagrams we obtain expressions that are proportional to
the internal baryon propagator. The denominator of the propagator is either
$ v \cdot p  = m_i - \mnod $ or $ v \cdot q = E_j - \mnod $. The values
of $ m_i $, the physical mass of the decaying baryon, and $ E_j $, the 
relativistic energy of the outgoing baryon, are fixed from experiment, 
since we are in the rest frame of the heavy baryon. On the other side,
$ \mnod $ must be predicted from theory, \cite{BM}. But this
quantity is not well known. The internal baryon propagator is 
of chiral order $ {\cal O}(p^{-2}) $ and very sensitive to modifications
in $ \mnod $. Different values for $ \mnod $ 
alter the results
for the p-waves significantly. In order to make our results more stable 
we replace 
$ \mnod $ by the physical mass of the internal baryon.
The remainder of the self-energy diagrams
of the internal baryon, which include the off-shell momentum and do not
directly contribute
to the mass, are considered only to the order we are working. 
In \cite{Spr} only a part of this remainder  has been considered.
The explicit forms of the 
coefficients in eq.(\ref{pwa}) can be found in app.~\ref{app:d}.

\subsection{Renormalization}
The loop contributions to the decays are, of course, divergent
and we must renormalize 
s- and p-waves separately. We start with the s-waves.
The corresponding loop diagrams are shown in fig.~2.
In order to calculate them we use dimensional regularization.
The mass dependent divergences can then be absorbed by the $h_i$ terms
\beq \label{divd}
h_i = h_i^r(\mu) + \frac{L}{24 \, F_\pi^2} \, \Gamma_i
\eeq
with $\mu$ being the scale of dimensional regularization and
\begin{equation}
L = \frac{\mu^{d-4}}{16 \pi^2} \biggl\lbrace \frac{1}{d-4} -
\frac{1}{2}[\ln (4 \pi) +1 - \gamma_E] \biggr\rbrace
\label{L}
\end{equation}
with $\gamma_E = 0.5772215..$ being the Euler-Mascheroni constant. The
scale dependence of the $h_i^r (\mu)$ follows from
eq.(\ref{divd}):
\begin{equation}
h_i^r (\mu_2) = h_i^r (\mu_1) + \frac{\Gamma_i}{24 \, (4 \pi F_\pi)^2} 
\ln \frac{\mu_1}{\mu_2} \quad .
\end{equation}

In the following, we set $\mu =1$~GeV.
Renormalizing the s-wave amplitudes one obtains for $\Gamma_i$ 
\beqa
\Gamma_3  & = & 7d   + 2D^2 d + 18F^2d - 12 DFf  \no \\
\Gamma_5  & = & \frac{21}{2} f  + 3 DFd + \frac{21}{2}D^2 f  
                - \frac{27}{2} F^2 f \no \\
\Gamma_7  & = & \frac{21}{2} f  -  3 DFd + \frac{27}{2}D^2 f  
                + \frac{27}{2} F^2 f   \no \\
\Gamma_8  & = & 14d  - 22D^2 d + 18F^2d + 36 DFf     \no \\
\Gamma_{11}  & = & -10f   + 36 DFd - 34D^2 f  -  18F^2 f   \no \\
\Gamma_{12}  & = & 4d  + 12D^2 d - 36F^2d +72 DFf    
\eeqa
The diagrams ~2$\,$c - ~2$\,$e also involve momentum 
dependent divergences, which are 
quadratic in the energies of the external on-shell particles, leading to terms
proportional to $(v \cdot p)^2$, $v \cdot p\, v \cdot q  $ and
$(v \cdot q)^2$. 
In order to keep the result finite one has to add the counterterms
\beqa
&&
\Big( \, - 3 d + 2 D^2 d + 12 DFf -6F^2d \Big)\, \frac{L}{F_\pi^2} \,
   {\rm tr}\Big(\bar{B} \{ [ v \cdot D, [ v \cdot D,
      h_+ ]] , B \} \Big) \no \\
&&
+ \Big( \, - 3f - \frac{10}{3} D^2f +\frac{20}{3} DFd - 6 F^2f \Big)\, 
   \frac{L}{F_\pi^2} {\rm tr}\Big(\bar{B} [ [ v \cdot D, [ v \cdot D,
      h_+ ]] , B ] \Big)  \no \\
&&
+3 \,\Big( \, d -2 D^2 d - 12 DFf + 6F^2d\Big)  \, \frac{L}{F_\pi^2} \,\,
{\rm tr}\Big([ v \cdot D,\bar{B}] \{  
      h_+ , [ v \cdot D,B] \} \Big) \no \\
&&
+3 \, \Big( \, f + \frac{10}{3} D^2f -\frac{20}{3} DFd + 6 F^2f \Big)\, \,
     \frac{L}{F_\pi^2} {\rm tr}\Big([ v \cdot D,\bar{B}] [  
      h_+ , [ v \cdot D,B] ] \Big) \no \\
&&
+ \frac{1}{2} f \, \frac{L}{F_\pi^2} \ci \, \Big\{
    {\rm tr}\Big(  \bar{B} [h_+ ,[ v \cdot u , [ v \cdot D,B]]] \Big)
- {\rm tr}\Big( [ v \cdot D,\bar{B}]  [ v \cdot u ,[h_+ ,B]]\Big) \Big\}\no \\
&&
- \frac{1}{2} f \, \frac{L}{F_\pi^2} \ci \, \Big\{
    {\rm tr}\Big( [ v \cdot D, \bar{B}] [h_+ ,[ v \cdot u , B]] \Big)
- {\rm tr}\Big( \bar{B}[ v \cdot u ,[h_+ ,[ v \cdot D , B]]]\Big) \Big\}\no \\
&&
+ \frac{2}{3} d \, \frac{L}{F_\pi^2} \ci \, \Big\{
    {\rm tr}\Big(  \bar{B} [h_+ ,\{ v \cdot u , [ v \cdot D,B]\}] \Big)
- {\rm tr}\Big( [ v \cdot D,\bar{B}] \{v \cdot u ,[h_+ ,B]\}\Big) \Big\}\no \\
&&
- \frac{2}{3} d \, \frac{L}{F_\pi^2} \ci \, \Big\{
    {\rm tr}\Big( [ v \cdot D, \bar{B}] [h_+ ,\{ v \cdot u , B\} ] \Big)
- {\rm tr}\Big(\bar{B}\{ v \cdot u ,[h_+ ,[v \cdot D , B]]\}\Big) \Big\}\no \\
&&
+ \frac{2}{3} d \, \frac{L}{F_\pi^2} \ci \, \Big\{
    {\rm tr}\Big(  \bar{B} \{h_+ ,[ v \cdot u , [ v \cdot D,B]]\} \Big)
- {\rm tr}\Big( [ v \cdot D,\bar{B}][ v \cdot u ,\{h_+ ,B\}]\Big) \Big\}\no \\
&&
- \frac{2}{3} d \, \frac{L}{F_\pi^2} \ci \, \Big\{
    {\rm tr}\Big( [ v \cdot D, \bar{B}] \{h_+ ,[ v \cdot u , B]\} \Big)
- {\rm tr}\Big( \bar{B}[ v \cdot u ,\{h_+ ,[v \cdot D , B]\}]\Big) \Big\}\no \\
&&
+ 3 f \, \frac{L}{F_\pi^2} \ci \, \Big\{
    {\rm tr}\Big(\bar{B} h_+ \Big) {\rm tr}\Big(v \cdot u [v \cdot D , B]\Big)
-{\rm tr}\Big([v \cdot D , \bar{B}]v\cdot u \Big) {\rm tr}\Big(h_+ B\Big)
Big\} \no \\
&&
- 3 f \, \frac{L}{F_\pi^2} \ci \, \Big\{
    {\rm tr}\Big([v \cdot D , \bar{B}] h_+ \Big) {\rm tr}\Big(v \cdot u B\Big)
-{\rm tr}\Big(\bar{B}v\cdot u \Big) {\rm tr}\Big(h_+[v \cdot D ,  B]\Big)Big\} 
\eeqa
This completes the renormalization of the s-waves.

Some of the above mentioned counterterms also contribute to the 
renormalization of the p-waves.
But in addition one has to include further higher order counterterms 
of the weak Lagrangian and also counterterms from the 
strong sector. To cancel the divergences arising
in the calculation of the p-wave amplitudes one has the prescription
\beq
H_i = H_i^r(\mu) + \frac{L}{48 \, F_\pi^2} \, \Gamma_i^{\prime}
\eeq
with the $H_i$ defined in eq.(\ref{eq:b17}) and
\beqa
\Gamma_4^{\prime}  & = & 3 D  - \frac{2}{3} D^3 -2 DF^2  \q , \q
\Gamma_5^{\prime}   =  \frac{9}{2} F  - \frac{7}{2}D^2 F 
                - \frac{9}{2} F^3  \no \\
\Gamma_6^{\prime}  & = & \frac{9}{2} F  - \frac{9}{2}D^2 F  
                + \frac{9}{2} F^3   \q , \q
\Gamma_{7}^{\prime}   =  6 D  + \frac{22}{3} D^3 - 18 DF^2    \no \\
\Gamma_{8}^{\prime}  & = & 6 F  - \frac{2}{3} D^2 F + 6 F^3   \q , \q 
\Gamma_{9}^{\prime}  =  12 D   - 4D^3  -  12D F^2  
\eeqa
For the momentum dependent divergences one has to include the terms
\beqa
&&
3 \, Dd \, \frac{L}{F_\pi^2} \ci \, \Big\{
    {\rm tr}\Big( \bar{B} S_{\mu} \{ u^{\mu} , \{ h_+ ,[v \cdot D , B]\}\}\Big)
- {\rm tr}\Big( [v \cdot D , \bar{B}] S_{\mu} \{h_+,\{u^{\mu} ,B\}\}\Big)
\Big\} \no \\
&+&
\Big(\,3 \, Df \, -\frac{5}{2}D^2 Fd \, - \frac{3}{2} F^3d 
     \, - \frac{1}{6} D^3f + \frac{3}{2} DF^2f \Big) \no \\
&& \qq \qq \times \, \, 
\frac{L}{F_\pi^2} \ci \, \Big\{
    {\rm tr}\Big( \bar{B} S_{\mu} \{ u^{\mu} , [ h_+ , [v \cdot D , B]]\}\Big)
- {\rm tr}\Big( [v \cdot D , \bar{B}] S_{\mu} [h_+,\{u^{\mu} , B\}]\Big)
\Big\} \no \\
&+&
\Big(\, -\frac{5}{2}D^2 Fd \, - \frac{3}{2} F^3d 
     \, - \frac{1}{6} D^3f + \frac{3}{2} DF^2f \Big) \no \\
&& \qq \qq \times \, \, 
\frac{L}{F_\pi^2} \ci \, \Big\{
    {\rm tr}\Big( \bar{B} S_{\mu} \{h_+  , [ u^{\mu} , [v \cdot D , B]]\}\Big)
- {\rm tr}\Big( [v \cdot D , \bar{B}] S_{\mu} [u^{\mu},\{h_+ , B\}]\Big)
\Big\} \no \\
&+&
\Big(\,3 \, Fd \, -\frac{7}{6}D^2 Fd \, - \frac{3}{2} F^3d 
     \, - \frac{3}{2} D^3f + \frac{3}{2} DF^2f \Big) \no \\
&& \qq \qq \times \, \, 
\frac{L}{F_\pi^2} \ci \, \Big\{
    {\rm tr}\Big( \bar{B} S_{\mu} [ u^{\mu} , \{ h_+ , [v \cdot D , B]\}]\Big)
- {\rm tr}\Big( [v \cdot D , \bar{B}] S_{\mu} \{h_+,[u^{\mu} , B]\}\Big)
\Big\} \no \\
&+&
\Big(-\frac{7}{6}D^2 Fd \, - \frac{3}{2} F^3d 
     \, - \frac{3}{2} D^3f + \frac{3}{2} DF^2f \Big) \no \\
&& \qq \qq \times \, \, 
\frac{L}{F_\pi^2} \ci \, \Big\{
    {\rm tr}\Big( \bar{B} S_{\mu} [ h_+ , \{ u^{\mu}, [v \cdot D , B]\}]\Big)
- {\rm tr}\Big( [v \cdot D , \bar{B}] S_{\mu} \{u^{\mu},[h_+ , B]\}\Big)
\Big\} \no \\
&+&
\Big(\,3 \, Ff \, -\frac{1}{9}D^3d \, + \frac{7}{3} DF^2d 
     \, + \frac{7}{3} D^2Ff - F^3f \Big) \no \\
&& \qq \qq \times \, \, 
\frac{L}{F_\pi^2} \ci \, \Big\{
    {\rm tr}\Big( \bar{B} S_{\mu} [ u^{\mu} , [ h_+ , [v \cdot D , B]]]\Big)
- {\rm tr}\Big( [v \cdot D , \bar{B}] S_{\mu} [h_+,[u^{\mu} , B]]\Big)
\Big\} \no \\
&+&
\Big(-\frac{1}{9}D^3d \, + \frac{7}{3} DF^2d 
     \, + \frac{7}{3} D^2Ff - F^3f \Big) \no \\
&& \qq \qq \times \, \, 
\frac{L}{F_\pi^2} \ci \, \Big\{
    {\rm tr}\Big( \bar{B} S_{\mu} [h_+ , [ u^{\mu}  , [v \cdot D , B]]]\Big)
- {\rm tr}\Big( [v \cdot D , \bar{B}] S_{\mu} [u^{\mu} ,[h_+, B]]\Big)
\Big\} \no \\
&+&
\Big( \, -4 \, Dd \, -\frac{2}{3}D^3d \, + 6 DF^2d 
     \, + 6 D^2Ff - 6 F^3f \Big) \no \\
&& \qq \qq \times \, \, 
\frac{L}{F_\pi^2} \ci \, \Big\{
 {\rm tr}\Big( \bar{B}u^{\mu}\Big) S_{\mu}{\rm tr}\Big( h_+[v \cdot D , B]\Big)
-{\rm tr}\Big( [v \cdot D ,\bar{B}] h_+\Big) S_{\mu}{\rm tr}\Big(u^{\mu}B\Big)
\Big\} \no \\
&+&
\Big( \, -\frac{2}{3}D^3d \, + 6 DF^2d 
     \, + 6 D^2Ff - 6 F^3f \Big) \no \\
&& \qq \qq \times \, \, 
\frac{L}{F_\pi^2} \ci \, \Big\{
 {\rm tr}\Big( \bar{B}h_+\Big) S_{\mu}{\rm tr}\Big( u^{\mu}[v \cdot D , B]\Big)
-{\rm tr}\Big( [v \cdot D ,\bar{B}]u^{\mu} \Big) S_{\mu}{\rm tr}\Big(h_+B\Big)
\Big\} \no \\
&+&
\Big(\,\frac{1}{3}D^3d \, - \frac{5}{3} DF^2d 
     \, - \frac{5}{3} D^2Ff + 3  F^3f \Big) \no \\
&& \qq \qq \times \, \, 
\frac{1}{3}\frac{L}{F_\pi^2} \ci \, \Big\{
    {\rm tr}\Big( \bar{B} S_{\mu} [ [v \cdot D , u^{\mu}] , [ h_+ , B]]\Big)
- {\rm tr}\Big( \bar{B} S_{\mu} [h_+,[[v \cdot D , u^{\mu}] , B]]\Big)
\Big\} \no \\
&+&
\Big(\, -\frac{13}{3}D^2 Fd \, + 3 F^3d 
     \, -  D^3f - 3 DF^2f \Big) \no \\
&& \qq \qq \times \, \, 
\frac{1}{3} \frac{L}{F_\pi^2} \ci \, \Big\{
    {\rm tr}\Big( \bar{B} S_{\mu} [ [v \cdot D , u^{\mu}] , \{ h_+ , B\}]\Big)
- {\rm tr}\Big( \bar{B} S_{\mu} \{h_+,[[v \cdot D , u^{\mu}] , B]\}\Big)
\Big\} \no \\
&+&
\frac{16}{27}D^3d \,\, \frac{L}{F_\pi^2} \ci \, \Big\{
 {\rm tr}\Big( \bar{B}[v \cdot D ,u^{\mu}]\Big) S_{\mu}{\rm tr}\Big( h_+ B\Big)
-{\rm tr}\Big( \bar{B} h_+\Big) S_{\mu}{\rm tr}\Big([v \cdot D ,u^{\mu}]B\Big)
\Big\} \no \\
&+&
\frac{2}{9} \, \frac{L}{F_\pi^2} \, \Big( -5D^2F + 9F^3 \Big) \, 
{\rm tr}\Big( \bar{B} S_{\mu} [[v \cdot D ,[v \cdot D ,u^{\mu}]] ,B]\Big)\no \\
&-&
\frac{2}{3} \, \frac{L}{F_\pi^2} \, \Big( D^3 + 3F^2D \Big) \, 
{\rm tr}\Big( \bar{B} S_{\mu}\{[v \cdot D ,[v \cdot D ,u^{\mu}]],B\}\Big)\no \\
&+&
\frac{2}{3} \, \frac{L}{F_\pi^2} \, \Big( 5D^2F - 9F^3 \Big) \, 
{\rm tr}\Big([v \cdot D , \bar{B}] S_{\mu} [u^{\mu} ,[v \cdot D ,B]]\Big)\no \\
&+&
2 \, \frac{L}{F_\pi^2} \, \Big(D^3 + 3F^2D  \Big) \, 
{\rm tr}\Big([v \cdot D ,\bar{B}] S_{\mu}\{u^{\mu} ,[v \cdot D ,B]\}\Big)\no \\
&+& \ci \frac{3 \, L}{4\, F_{\pi}^2} (D^2 -3F^2) 
\, {\rm tr}(\bar{B} \{ \chi_+, [ v \cdot D, B]\})
 - \ci \frac{5 \, L}{2\, F_{\pi}^2} DF \, 
           {\rm tr}(\bar{B} [ \chi_+, [ v \cdot D, B]]) \nonumber \\
 &-& \ci \frac{3\, L}{2\, F_{\pi}^2} 
       (\frac{13}{9}D^2 +F^2) \, {\rm tr}(\bar{B}  [ v \cdot D, B])
{\rm tr}( \chi_+) 
\eeqa
This completes the renormalization of the p-waves.
Note, that the renormalization of the pure strong sector is in agreement with 
\cite{MM}.

\section{Results and discussion} \label{sec:dis}
In this section we discuss the numerical values of the LECs and the 
fit to experiment. There exist eight independent
experimental numbers, {\it i.e.} s- and 
p-wave amplitudes for the four decays 
$\Sigma^+  \rightarrow n \, \pi^+ \, , \, 
\Sigma^- \rightarrow n \, \pi^- \, , \, 
\Lambda \rightarrow p \, \pi^- \, \mbox{and} \,
\Xi^- \rightarrow  \Lambda \, \pi^- $,
which are not related by isospin. The central values for our parameters
are $ F_{\pi} = 93 $MeV, $ D= 0.75$, $F=0.50$ and we set $ \mu = 1.0$GeV.
For $\mnod$, the octet baryon mass in the chiral limit, we use 
$ \mnod = 767 $MeV \cite{BM}. 
For the various mesonic LECs $L_i^r
(\mu)$, we use the central values taken from the compilation of
Bijnens et al. in ref.\cite{daphne}. 

Initially, we neglect the counterterms $g_{11}$ to $g_{16}$ from the weak
Lagrangian ${\cal L}_{\phi B}^{W \, (1)}$ since they are presumed to be
absent within the resonance saturation picture. (Note, that $g_{18}$ does
not contribute to the decay amplitudes.)
This leaves us with just the coupling constants
$d,f$ and
$h_3,h_5, h_7, h_8$ after the four LECs $h_1,h_2, h_{11}, h_{12}$ 
have been absorbed into $d$ and $f$. 
A simple least-squares fit to the decay amplitudes turns out, however, to be 
very unsatisfactory. 
Although the s-waves can be well fit, 
there exist large discrepancies between the results
and the experimental values for the p-wave decay amplitudes. 
A simultaneous fit of s- and p-waves is impossible -- there is also
no recognizable convergence in the chiral expansions and the results 
are not realistic, so we do not present them here.
One can disentangle
$d,f$ and the four LECs $h_{1}, h_{2}, h_{11}, h_{12}$ in order to
perform a better fit. But the eight LECs $h_i$ appear only in four
different combinations in the expressions for the decay amplitudes, so that
a similar least--squares fit has to be performed yielding
the same result.
This leads us to the conclusion that the estimation
of the LECs via the resonance saturation principle is very unsatisfactory.

In order to obtain a good fit to the decay amplitudes
it is necessary to go beyond the
resonance estimate hypothesis. As the simplest such possibility we take the
terms $g_{11}$ to $g_{16}$ into account.
In this case, we have the ten unknown weak LECs $d,f$ and
$h_3,h_5, h_7, h_8,g_{11},g_{13},g_{15},g_{16}$. 
It is, of course, then possible to fit the eight independent
decay amplitudes exactly in many different ways.
We will perform the fit as described below, delivering reasonable
results.
Note, that the coupling constants 
$d,f$ and the other LECs should not be treated on the same level,
since the 
former contribute at lowest order, whereas the latter constitute only  
higher order corrections.

The p-waves are sensitive to relatively small 
changes in the parameters $d$ and $f$,
since the pole diagrams contributing to the p-wave decays all involve
cancellations between two or more diagrams with opposite signs 
\cite{Jen},
yielding a final
result smaller than the individual components. This suggests that the
higher order terms that have been neglected in former papers, 
\cite{Bij,Jen, CG}, will
play a more important role for the p-waves than for the s-waves.
In light of this sensitivity of the p-wave amplitudes, we elect to first
perform a least-squares fit to just the s-waves for $d$ and $f$ by using
only the tree result.
After that we perform a fit to the complete expressions to 
second chiral order of the s-waves by using the LECs
$h_3,h_5, h_7$ and $ h_8$, but keeping $d$ and $f$ fixed.
The decay amplitude ${\cal A}_{\Sigma^+ n}^{(s)}$ cannot be fit
since it derives no contribution from the counterterms considered here
and has a nonvanishing experimental value.
Thus we have to impose an additional constraint on the $h_i$ which we 
arbitrarily choose
to be $h_5 = h_7$. (As it turns out replacing this constraint by a
different realistic one does not alter the results significantly.
In the following we will therefore work with $h_5 = h_7$.)
Finally, we are able to fit the p-waves exactly by using the
counterterms $g_{11},g_{13},g_{15},g_{16}$, which contribute only 
to the p-waves, and keeping the other LECs fixed.

A consistent picture emerges. 
The chiral expansions of the decay 
amplitudes read in units of $10^{-7}$
\beqa
{\cal A}_{\Sigma^+ n}^{(s)} & = & 0.0 +  0.0 + 0.0 =  0.0 \q ,\q
{\cal A}_{\Sigma^+ n}^{(p)}   =   1.47 + 96.1 - 53.2 =  44.4 \q ,\q \no \\
{\cal A}_{\Sigma^- n}^{(s)} & = & 4.37 + 1.29 - 1.39 = 4.27 \q ,\q
{\cal A}_{\Sigma^- n}^{(p)}   =   7.37 - 3.50 - 5.39 = -1.52 \q ,\q \no \\
{\cal A}_{\Lambda p}^{(s)} & = &   3.33 + 2.08 - 2.16 = 3.25 \q ,\q
{\cal A}_{\Lambda p}^{(p)}   =   -25.9 - 2.96 + 52.26 = 23.4 \q ,\q \no \\
{\cal A}_{\Xi^- \Lambda}^{(s)} & = & -4.34 - 1.74 + 1.57 = -4.51 \q ,\q
{\cal A}_{\Xi^- \Lambda}^{(p)}   =   7.37 + 10.16 - 2.73 = 14.8 \q ,
\eeqa
where the first number is the lowest order contribution, the second
number contains the nonanalytic pieces to the decay amplitudes and
the contributions of the higher order counterterms are
summarized in the third number. 
We observe that the s--wave results show reasonable convergence
of the chiral expansion. However,
there are large 
contributions in the higher orders for the p-waves, especially
for ${\cal A}_{\Sigma^+ n}^{(p)} $ and $ {\cal A}_{\Lambda p}^{(p)}$.
The experimental values for the decay amplitudes can be found in table~1.
The numerical values for the LECs are 
presented in table~2.

It is interesting to note that
the Lee-Sugawara relation \cite{LS}, which is a prediction of  $SU(3)$ 
symmetry, reads
\beq
{\cal A}_{\Lambda p} + 2 {\cal A}_{\Xi^- \Lambda}
+ \sqrt{\frac{3}{2}} \Big( {\cal A}_{\Sigma^- n} 
                - {\cal A}_{\Sigma^+ n} \Big)  = 0
\eeq
and is exactly fulfilled for the s-waves to lowest order. Adding the
higher order contributions we find for s-waves (in units of $10^{-7}$)
\beq
{\cal A}_{\Lambda p}^{(s)} + 2 {\cal A}_{\Xi^- \Lambda}^{(s)}
+ \sqrt{\frac{3}{2}} \Big( {\cal A}_{\Sigma^- n}^{(s)}  
                - {\cal A}_{\Sigma^+ n}^{(s)}  \Big)  = 
                  0  - 0.54 = -0.54 \qq .
\eeq
After disentangling the various contributions, we obtain for p-waves
 \beq
{\cal A}_{\Lambda p}^{(p)} + 2 {\cal A}_{\Xi^- \Lambda}^{(p)}
+ \sqrt{\frac{3}{2}} \Big( {\cal A}_{\Sigma^- n}^{(p)}  
                - {\cal A}_{\Sigma^+ n}^{(p)} \Big)   = 
                 -3.95 + 0.77 = - 3.18 \qq .
\eeq
The final results on the right side of the equations are much smaller
than the individual terms on the left-hand side. The pertinent experimental
values are $-0.70 \times 10^{-7}$ and $-3.18 \times 10^{-7}$ 
for s- and p-waves, respectively.

\subsection{Theoretical uncertainties}
In the previous section we gave the results for the central values
of the  parameters $F_{\pi}, D, F, \mu$ and $\mnod$ . 
Here, we will discuss the spread of
the results due to uncertainties related to these numbers. 

Consider first the dependence on the octet baryon mass in the chiral limit
$\mnod$. In order to understand the uncertainty 
in this variable we choose the nucleon mass,
$\mnod = 940$MeV. The variations in the fitted numerical values
of the LECs can be found in table~2. 
In our results
$\mnod$ is contained only in the relativistic corrections and a 
variation in $\mnod$ does not alter the results considerably.
Next, we consider a variation in the coupling constants $D$ and $F$.
For comparison with our central values we use 
$D= 0.85 \pm 0.06$, $F= 0.52 \pm  0.04$
given by Luty and White \cite{luwh}.
Finally, we alter the scale of dimensional regularization $\mu$.
This dependence is introduced since we neglect some of  the LECs 
of the entire Lagrangian
and would disappear once all LECs could be 
determined from data. In table~2
we show the results for the range
$0.8 \, {\rm GeV} \le \mu \le 1.2\,$GeV,  
for the central values of $F_\pi$, $F,D$ and $\mnod$.
We therefore assign the follwing theoretical uncertainties
to the results of the LECs 
$h_3,h_5, h_7, h_8$
and $g_{11},g_{13},g_{15},g_{16}$ after setting $h_5 = h_7$.
\beqa
h_3 &=&  0.03 \pm 0.06 \q , \q h_5  = 0.10 \pm 0.06 \no \\
h_7 &=& 0.10 \pm 0.06 \q , \q h_8  =  0.08 \pm 0.11  \no \\
g_{11} &=&  - 0.48 \pm 0.04\q , \q  g_{13} =  - 0.20 \pm 0.08 \no \\
g_{15} &=&  0.44 \pm 0.07\q , \q  g_{16} = - 3.76 \pm 0.60 
\eeqa
The numbers are given in units of $10^{-7}$GeV$^{0}$ and
$10^{-7}$GeV$^{-1}$ for the $g_i$ and $h_i$, respectively.
In our scheme of fitting LECs to experiment, the values
for $d$ and $f$ do not change when varying the above mentioned
parameters. We therefore cannot quote errorbars for these couplings.
Note also, that the LEC $g_{16}$ has a much larger value than the other LECs.
This is due to the large nonanalytic correction for the decay
$\Sigma^+ \rightarrow n \pi^+$ which is mainly compensated by $g_{16}$.
The counterterm $g_{16}$ contributes only to this decay.

The uncertainties in the low-energy constants 
do not include the possible effects of higher
orders, which can only be assessed if one performs a multi-loop calculation.
This, however, goes beyond the scope of the present paper.

\section{Summary and conclusions}
In this paper, we have considered the decay amplitudes for 
the non-leptonic hyperon decays, to linear (quadratic) order
in the quark (Goldstone boson)
masses, in the framework of heavy baryon chiral perturbation
theory. The key results of this investigation can be summarized
as follows:
\begin{enumerate}
\item[$\circ$]
We have constructed the most general weak effective Lagrangian
to ${\cal O}(p^2)$ in the small parameter $p$ (external momentum
or meson mass) and to ${\cal O}(p^3)$
for the strong effective Lagrangian
necessary to investigate the decay amplitudes.
For the weak Lagrangian we have introduced two independent combinations
of the spurion field $h$, that transform like mass fields,
and have also included kinematical $1/  \mnod$ and $1/  \mnod^2$ corrections.

\item[$\circ$]
We are unable to fix the weak 
LECs strictly from experiment even if we were to
resort to large $N_c$ arguments. For the strong Lagrangian, 
$D= 0.75$ and $F=0.50$ give a satisfactory fit to semileptonic
hyperon decay data, and
we therefore neglect the higher order 
contributions to the axial-vector couplings $D$ and $F$ in this Lagrangian.
For the weak Lagrangian we first attempted to use the exchange of the $\Delta$
resonance as an indication which LECs are important.
There exist then two LECs at lowest order ${\cal O}(p^0)$ -- $d$ and $f$ -- 
and eight at order ${\cal O}(p^2)$ -- 
$h_{1,2,3,5,7,8,11,12}$.
Four of the latter amount to quark mass renormalizations of $d$ and $f$
and can be absorbed after an appropriate redefinition of
these coupling constants.
But it turns out that
there exist large discrepancies between the results of such a fit
and the experimental values for the decay amplitudes -- there is
no recognizable convergence in the chiral expansions and the results 
are not realistic. This seems to indicate
that we have neglected some significant LECs.
The obvious solution is to include the counterterms
from the next--to--leading order Lagrangian ${\cal L}_{\phi B}^{W \, (1)}$.
This was also suggested
in \cite{Neu} where 
a rough estimate of the LECs of the weak baryon Lagrangian
of order ${\cal O}(p)$ has been given using the weak deformation model. 
The author comes to the conclusion that one cannot understand
nonleptonic hyperon decays without such terms.
We agree with this assertion and conclude that one must include four new LECs 
$g_{11},g_{13},g_{15},g_{16}$ which contribute only to the p-waves.
In order to estimate the LECs, we first perform a 
least-squares fit to the s-waves for $d$ and $f$ using 
only the tree level result.
The reason for not including the p-waves in this fit is that
in this case the higher order corrections are much more significant
than for the s-waves due to cancellations between the pole diagrams.
For the higher order LECs $h_i$ a fit 
is then performed by applying
the complete expression for the s-waves.
The remaining LECs
$g_{11},g_{13},g_{15},g_{16}$ are then fitted by applying
the entire expressions for the p-waves.
We achieve an excellent fit
to the experimental values of the decay amplitudes.
The chiral expansions for the s-waves are reasonably well behaved
whereas for the p-waves we find significant higher order contributions,
especially for ${\cal A}_{\Sigma^+ n}^{(p)}$ and
${\cal A}_{\Xi^- \Lambda}^{(p)}$. 

\item[$\circ$]
A possible approach to improving the convergence of the chiral
expansion might be to include the decuplet
as explicit degrees of freedom.
A first step towards this direction has already been 
undertaken in \cite{Jen, Spr} but only the leading non-analytic pieces 
from the loops were retained. In order to get the full picture
one has to account for
all counterterms. This would avoid the uncertainty in estimating
the LECs via the resonance saturation scheme, but on the other
hand introduce new unknown coupling constants.
Such a calculation, however, is far beyond the scope of this work.

\end{enumerate}

\section*{Acknowledgements}
We thank Joachim Kambor and Guido M{\"u}ller for discussions and useful 
comments. We are grateful to Ulf-G. Mei{\ss}ner for conversations and 
reading the manuscript. We also thank Gerhard Ecker for several comments.

\appendix 
\def\theequation{\Alph{section}.\arabic{equation}}
\setcounter{equation}{0}
\section{Construction principles for the relativistic Lagrangian} \label{app.a}
In this appendix we present some construction principles for
the most general Lagrangian in the relativistic 
formulation which is invariant 
under $ CPS $ and chiral transformations. 
The transformation $S$
interchanges down and strange quarks in the Lagrangian.
We will work in the $CP$ limit so that all LECs are real.
Note, that $C$ and $P$ invariance 
are not required separately.
The weak interactions start at zeroth chiral order whereas the strong 
interactions begin at first order. It immediately follows from chiral 
counting that to the order we are working one needs the weak Lagrangian 
up-to-and-including second order and to third order for the strong sector. 
For this purpose, it is convenient to use the combination
\beq
h_+ = u^{\dagger} h u +  u^{\dagger} h^{\dagger} u  \qquad , \qquad
\eeq
with $h^{a}_{b} = \delta^{a}_{2} \delta^{3}_{b}$ the weak transition matrix.
$h_+$ transforms as matter field.
Under $CP$ transformations the fields behave like follows
\beqa
B \q & \rightarrow & \q \gamma_0 C \bar{B}^{T} \q , \q
\bar{B} \q \rightarrow  \q B^{T} C \gamma_0  \q , \q
u^{\mu} \q \rightarrow  \q - u_{\mu}^{T} \q , \q \no \\
h_+ \q & \rightarrow &  \q h_+^{T} \qq , \qq
D^{\mu} \q \rightarrow  \q - D_{\mu}^{T} \q , \q \no \\
\chi_+ \q & \rightarrow & \q \chi_+^{T} \qq , \qq
\chi_- \q \rightarrow  \q - \chi_-^{T} \qq ,
\eeqa
where $C$  is the usual charge conjugation matrix.
There are some relations which can be used to reduce the number of 
independent terms in the Lagrangian. First there is the equation 
of motion (eom) 
for the baryons, which to lowest order it reads
\beq
i \gamma_{\mu} [ D^{\mu}, B] - \mnod B  = 0
\eeq
with an analogous relation for $\bar{B}$. Terms of higher orders in the eom 
are neglected here since they can be absorbed by appropriate counterterms.
Using the eom one can reduce the number of derivatives acting on the baryon 
field -- {\it e.g.} it turns out that the terms of the form 
$ \q
\mbox{tr} \Big( \bar{B} \sigma_{\mu \nu} \{A^{\nu} , [ D^{\mu}, B] \} \Big)
\q $
can be neglected after decomposing the $\sigma_{\mu \nu}$ in terms of
$\gamma$  matrices.
Here, $\, A^{\nu} \,$ denotes any combination of fields and there 
are analogous terms where the anticommutator 
is replaced by the commutator.
Another relation is
\beq
\tr \Big( \bar{B} \gamma_{\mu} ( A^{\mu \nu} , [ D_{\nu}, B] ) \Big)
\simeq
\tr \Big( \bar{B} \gamma_{\nu} ( A^{\mu \nu} , [D_{\mu}, B] ) \Big) \qq ,\no
\eeq
where $ \simeq $ stands for the equality up to terms of higher order.

Second, there are the Cayley-Hamilton identities.
For two traceless $3 \times 3$ matrices  $ A_1$ and  $ A_2$ the pertinent 
Cayley-Hamilton identity reads
\beqa
& & \tr  \Big( \bar{B} \{A_1, \{ A_2 , B \} \} \Big)
+\frac{1}{2} \tr \Big( \bar{B} \{A_2, \{ A_1 , B \} \} \Big)
+\frac{1}{2} \tr \Big( \bar{B} [A_2, [ A_1 , B ] ] \Big) \no \\
& = &  \tr \Big( \bar{B} B \Big)  \tr \Big( A_1  A_2 \Big)
+ \tr \Big( \bar{B} A_1 \Big)  \tr \Big(  A_2 B \Big)
+ \tr \Big( \bar{B} A_2 \Big)  \tr \Big(  A_1 B \Big) \q .
\eeqa
For the case with only $ A_1 $ traceless this identity becomes
\beqa
& & \tr  \Big( \bar{B} \{A_1, \{ A_2 , B \} \} \Big)
+\frac{1}{2} \tr \Big( \bar{B} \{A_2, \{ A_1 , B \} \} \Big)
+\frac{1}{2} \tr \Big( \bar{B} [A_2, [ A_1 , B ] ] \Big) \no \\
& = &  \tr \Big( \bar{B} B \Big)  \tr \Big( A_1  A_2 \Big)
+ \tr \Big( \bar{B} A_1 \Big)  \tr \Big(  A_2 B \Big)
+ \tr \Big( \bar{B} A_2 \Big)  \tr \Big(  A_1 B \Big) +
\tr \Big( \bar{B}  \{ A_1 , B \} \Big)  \tr \Big( A_2\Big) \q .
\eeqa
and these are the only Cayley-Hamilton identities we need here.

The total Lagrangian can be decomposed as follows
\beq
{\cal L}_{\mbox{eff}}  =  {\cal L}_{\phi B} + 
        {\cal L}_{\phi B}^W     + {\cal L}_{\phi}+{\cal L}_{\phi }^W  
\eeq
with the strong and weak mesonic Lagrangians $ {\cal L}_{\phi} $ 
and ${\cal L}_{\phi }^W $, respectively, as given in \cite{GL1}and 
eq.~(\ref{weak}).\\
For the weak meson-baryon Lagrangian one gets
\beq
{\cal L}_{\phi B}^W   = \:
{\cal L}_{\phi B}^{W \, (0)} \, +\,
{\cal L}_{\phi B}^{W \, (1)}  \, +\,
{\cal L}_{\phi B}^{W \, (2)}   \qq , 
\eeq
where the superscript denotes the chiral order. Since we will work in the
heavy baryon formalism, we do not list the whole Lagrangian explicitely. The
pertinent heavy baryon Lagrangian which one gets after integrating out 
the heavy degrees of freedom is shown in the next appendix.

Finally, the strong meson-baryon Lagrangian reads
\beq
{\cal L}_{\phi B} =
 {\cal L}_{\phi B}^{(1)} + {\cal L}_{\phi B}^{(2)} 
                  + {\cal L}_{\phi B}^{(3)} 
\eeq
with $ {\cal L}_{\phi B}^{(1)} $ the usual meson-baryon Lagrangian 
to lowest order . Here $ {\cal L}_{\phi B}^{(2)} $ does not contribute to the 
order we are working while $ {\cal L}_{\phi B}^{(3)} $ decomposes into
\beq
{\cal L}_{\phi B}^{(3)} =
{\cal L}_{\phi B}^{(3,br)} + \sum_i \, H_i \, O_i^{(3)}
\eeq
where $  {\cal L}_{\phi B}^{(3,br)} $ explicitely breaks the chiral
symmetry and the $ O_i^{(3)} $ denote monomials in the fields of 
chiral order three.

\def\theequation{\Alph{section}.\arabic{equation}}
\setcounter{section}{1}
\setcounter{equation}{0}
\section{The non-relativistic Lagrangian} \label{app.b}
The purpose of this appendix is to present the effective Lagrangian in
the heavy baryon formalism. Starting from the relativistic Lagrangian of 
appendix~\ref{app.a} one integrates out the heavy degrees of freedom. 
To this end the
baryon field $ B$ is split into upper and lower components with fixed
four-velocity $ v $
\beqa
B_v  & = &  \e^{ i \mnod v \cdot x } \frac{1}{2} ( 1 +  v \!\! /) B \no \\
b_v  & = &  \e^{ i \mnod v \cdot x } \frac{1}{2} ( 1 -  v \!\! /) B 
\eeqa
In the heavy mass formulation the Dirac algebra simplifies considerably and any
Dirac bilinear can be expressed in terms of the four-velocity $ v_{\mu} $
and the spin-operator $ 2 S_{\mu} = \ci  \gamma_5 \sigma_{\mu \nu} v^{\nu} $.
The effective Lagrangian can then be derived by the path integrals.
In this formulation, the $ 1/ \!\! \mnod $ corrections are easily constructed.
This method is outlined e.g. in \cite{MRT} and \cite{BKKM} and will not 
be repeated here. We only state our result.
For the sake of simplicity we will omit the index $ v $ from the field 
$ B_v $ and the Lagrangian will be denoted by $ \cal L $ as in the relativistic
case.
The Lagrangian can be written as follows
\beq
{\cal L}_{\mbox{eff}}  =  {\cal L}_{\phi B} + 
        {\cal L}_{\phi B}^W    + {\cal L}_{\phi}
\eeq
with the same mesonic Lagrangian ${\cal L}_{\phi}$
as in the relativistic case and 
\beq
{\cal L}_{\phi B}^W   = \:
{\cal L}_{\phi B}^{W \, (0)}  \, +\,
{\cal L}_{\phi B}^{W \, (1)}  \, +\,
{\cal L}_{\phi B}^{W \, (2)}    
\eeq
\beq
{\cal L}_{\phi B}^{W \, (0)}   =  \:
d \, \tr \Big( \bar{B}  \{ h_+ , B\} \Big) + \:
f \, \tr \Big( \bar{B}  [ h_+ , B ] \Big)
\eeq
\beq
{\cal L}_{\phi B}^{W \, (1)}     = 
\sum_{i} \, g_i O_i^{(1)} 
\eeq
with the $ O_i^{(1)} $ monomials in the fields of chiral order one.
The set of such terms is given by
\beqa
\sum_{i} \, g_i O_i^{(1)} & = &
g_3 \bigg\{ \tr \Big( \bar{B}  [ h_+ , [ v \cdot u, B] ] \Big) +
     \tr \Big( \bar{B}  [v \cdot u  , [h_+, B] ] \Big) \bigg\} \no \\
& + & 
g_5 \bigg\{ \tr \Big( \bar{B}  [ h_+ , \{v \cdot u , B\} ] \Big) +
  \tr \Big( \bar{B}  \{ v \cdot u , [h_+, B] \} \Big) \bigg\} \no \\
& + & 
g_7 \bigg\{ \tr \Big( \bar{B}  \{ h_+ , [v \cdot u, B] \} \Big) +
  \tr \Big( \bar{B}  [ v \cdot u , \{ h_+, B\} ] \Big) \bigg\} \no \\
& + & 
g_8 \bigg\{ \tr \Big( \bar{B} h_+ \Big)  \tr \Big( v \cdot u  B \Big)
+ \tr \Big(\bar{B}v \cdot u \Big) \tr \Big( h_+ B \Big)
\bigg\} \no \\ 
& + & 
g_{10} \tr \Big( \bar{B}  B \Big) \tr \Big( v \cdot u \,  h_+ \Big)\no \\
& + & 
2 g_{11} \bigg\{ \tr \Big( \bar{B} S_{\mu}  
                             [ h_+ , [u^{\mu}, B] ] \Big) +
     \tr \Big( \bar{B} S_{\mu}  
                         [ u^{\mu} , [h_+, B] ] \Big) \bigg\} \no \\
& + & 
2 g_{13} \bigg\{ \tr \Big( \bar{B} S_{\mu}  
                            [ h_+ , \{ u^{\mu}, B\} ] \Big) +
  \tr \Big( \bar{B} S_{\mu}  
                      \{ u^{\mu} , [h_+, B] \} \Big) \bigg\} \no \\
& + & 
2 g_{15} \bigg\{ \tr \Big( \bar{B} S_{\mu}  
                               \{ h_+ , [u^{\mu}, B] \} \Big) +
  \tr \Big( \bar{B} S_{\mu}  
                          [ u^{\mu} , \{ h_+, B\} ] \Big) \bigg\} \no \\
& + & 
2 g_{16} \bigg\{ \tr \Big( \bar{B} h_+ \Big) S_{\mu}  
                              \tr \Big( u^{\mu} B \Big)
+ \tr \Big(\bar{B} u^{\mu}\Big) S_{\mu}   
                                 \tr \Big( h_+ B \Big) \bigg\} \no \\ 
& + & 
2 g_{18} \tr \Big( \bar{B} S_{\mu}  B \Big) 
                             \tr \Big( u^{\mu} h_+ \Big)
\eeqa
In the next order appear explicit symmetry breaking terms besides the
relativistic corrections and double--derivative terms.
\beq
{\cal L}_{\phi B}^{W \, (2)}     = 
{\cal L}_{\phi B}^{W \, (2,br)}  \: +  \:
\sum_{i} \, h_i O_i^{(2)} \: +  \: 
{\cal L}_{\phi B}^{W \, (2,rc)}  
\eeq
\beqa
& &{\cal L}_{\phi B}^{W \, (2,br)}  \no \\ 
& = &
h_3 \bigg\{ \tr \Big( \bar{B}  [ h_+ , [ \chi_+, B] ] \Big)  
         +\tr \Big( \bar{B}  [ \chi_+, [ h_+ , B ]]  \Big) \bigg\} \no \\ 
& + & 
h_5 \bigg\{ \tr \Big( \bar{B}  [ h_+ , \{ \chi_+, B\} ] \Big)  
         +\tr \Big( \bar{B}  \{ \chi_+, [ h_+ , B ]\}  \Big) \bigg\} \no \\ 
& + & 
h_7 \bigg\{ \tr \Big( \bar{B}  \{ h_+ , [ \chi_+, B] \} \Big)  
         +\tr \Big( \bar{B}  [ \chi_+, \{ h_+ , B \} ]  \Big) \bigg\} \no \\ 
& + & 
h_8 \bigg\{ \tr \Big( \bar{B} h_+ \Big)  \tr \Big( \chi_+  B \Big)
+ \tr \Big(\bar{B} \chi_+ \Big) \tr \Big( h_+ B \Big)
\bigg\} 
\: + \: h_{10} \tr \Big( \bar{B}  B \Big) \tr \Big( \chi_+  h_+ \Big)\no \\
& + & 
h_{11} \tr \Big( \bar{B}  [ h_+ , B ] \Big) \tr \Big( \chi_+ \Big) +
h_{12} \tr \Big( \bar{B}  \{ h_+ , B ]\} \Big) \tr \Big( \chi_+ \Big) \no \\
& + & 
h_{13} \bigg\{ \tr \Big( \bar{B}  [ h_+ , [ \chi_-, B] ] \Big)  
         -\tr \Big( \bar{B}  [ \chi_-, [ h_+ , B ]]  \Big) \bigg\} \no \\ 
& + & 
h_{15} \bigg\{ \tr \Big( \bar{B}  [ h_+ , \{ \chi_-, B\} ] \Big)  
         -\tr \Big( \bar{B}  \{ \chi_-, [ h_+ , B ]\}  \Big) \bigg\} \no \\ 
& + & 
h_{18} \bigg\{ \tr \Big( \bar{B} h_+ \Big)  \tr \Big( \chi_-  B \Big)
- \tr \Big(\bar{B} \chi_- \Big) \tr \Big( h_+ B \Big)
\bigg\} 
\eeqa
\beqa
& & \sum_{i} \, h_i O_i^{(2)} \no \\ 
& = &
h_1 \, \tr \Big( \bar{B}  [ [ D_{\mu}, [D^{\mu}, h_+]] , B ] \Big) +
h_2 \, \tr \Big( \bar{B}  \{ [ D_{\mu}, [D^{\mu}, h_+]] , B \} \Big) \no \\ 
& + & 
\ci \, h_{23} \, \bigg\{ \tr \Big( \bar{B}  
[ h_+,[ [ D_{\mu}, u^{\mu}], B] ] \Big) 
- \tr \Big( \bar{B}  [ [ D_{\mu}, u^{\mu}],[ h_+ , B] ] \Big) \bigg\} \no \\ 
& + & 
\ci \,  h_{25} \,  \bigg\{ \tr \Big( \bar{B}
 [ h_+,\{ [ D_{\mu}, u^{\mu}], B\} ] \Big) 
- \tr \Big( \bar{B} \{ [ D_{\mu}, u^{\mu}],[ h_+ , B] \} \Big) \bigg\} \no \\ 
& + & 
\ci \, h_{28} 
 \, \bigg\{ \tr \Big( \bar{B} h_+ \Big)  
\tr \Big( [ D_{\mu}, u^{\mu}]  B \Big)
- \tr \Big(\bar{B} [ D_{\mu}, u^{\mu}] \Big) \tr \Big( h_+ B \Big)
\bigg\} \no \\ 
& + & 
2 \, h_{31} \, \epsilon_{\mu \nu \alpha \beta} \, v^{\alpha} \, 
\bigg\{ \tr \Big( \bar{B}  
               S^{\beta}      [ h_+,[ [ D^{\mu}, u^{\nu}], B] ] \Big) 
+ \tr \Big( \bar{B}  S^{\beta} 
               [ [ D^{\mu}, u^{\nu}],[ h_+ , B] ] \Big) \bigg\} \no \\ 
& + & 
2 \, h_{33} \, \epsilon_{\mu \nu \alpha \beta} \, v^{\alpha} \, 
\bigg\{ \tr \Big( \bar{B}  
        S^{\beta}             [ h_+,\{ [ D^{\mu}, u^{\nu}], B\} ] \Big) 
+ \tr \Big( \bar{B} S^{\beta} 
          \{ [ D^{\mu}, u^{\nu}],[ h_+ , B] \} \Big) \bigg\} \no \\ 
& + & 
2 \, h_{35} \, \epsilon_{\mu \nu \alpha \beta} \, v^{\alpha} \, 
\bigg\{\tr \Big( \bar{B}  
    S^{\beta}    \{ h_+,[ [ D^{\mu}, u^{\nu}], B] \} \Big) 
+ \tr \Big( \bar{B} S^{\beta} 
            [ [ D^{\mu}, u^{\nu}],\{ h_+ , B\} ]\Big)\bigg\} \no \\ 
& + & 
2 \, h_{36} \, \epsilon_{\mu \nu \alpha \beta} \, v^{\alpha} \, 
\bigg\{ \tr \Big( \bar{B} 
h_+ \Big)  S^{\beta} 
\tr \Big( [ D^{\mu}, u^{\nu}]  B \Big)
+ \tr \Big(\bar{B} [ D^{\mu}, u^{\nu}] \Big) S^{\beta} 
           \tr \Big( h_+ B \Big) \bigg\} \no \\ 
& + & 
2 \, h_{38} \, \epsilon_{\mu \nu \alpha \beta} \, v^{\alpha} \, 
\tr \Big( \bar{B} 
           S^{\beta}  B \Big) 
    \tr  \Big( [ D^{\mu}, u^{\nu}]  h_+ \Big) \no \\
& + & 
2 \, \ci \, h_{39} \, \bigg\{ \tr \Big( \bar{B}  S_{\mu} 
                     [ h_+,[ [ v \cdot D , u^{\mu}], B] ] \Big) 
- \tr \Big( \bar{B}  S_{\mu} 
               [ [ v \cdot D , u^{\mu}],[ h_+ , B] ] \Big) \bigg\} \no \\ 
& - & 
2 \, \ci \, \Big( \, h_{39} +  \mnod h_{55} \, \Big) \, 
\bigg\{ \tr \Big( \bar{B}  S_{\mu} 
                     [ h_+,[ [  D^{\mu} ,v \cdot u], B] ] \Big) 
- \tr \Big( \bar{B}  S_{\mu} 
               [ [ D^{\mu} , v \cdot u],[ h_+ , B] ] \Big) \bigg\} \no \\ 
& + & 
2 \, \ci \, h_{41} \, \bigg\{ \tr \Big( \bar{B}  S_{\mu} 
                     [ h_+,\{ [ v \cdot D , u^{\mu}], B\} ] \Big) 
- \tr \Big( \bar{B}  S_{\mu} 
               \{ [ v \cdot D , u^{\mu}],[ h_+ , B] \} \Big) \bigg\} \no \\ 
& - & 
2 \, \ci \, \Big( \, h_{41} +  \mnod h_{57} \, \Big) \, 
\bigg\{ \tr \Big( \bar{B}  S_{\mu} 
                     [ h_+,\{ [  D^{\mu} ,v \cdot u], B\} ] \Big) 
- \tr \Big( \bar{B}  S_{\mu} 
               \{ [ D^{\mu} , v \cdot u],[ h_+ , B] \} \Big) \bigg\} \no \\ 
& + & 
2 \, \ci \, h_{44} \, \bigg\{ \tr \Big( \bar{B} h_+ \Big)  S_{\mu}
\tr \Big( [ v \cdot D, u^{\mu}]  B \Big)
- \tr \Big(\bar{B} [ v \cdot D , u^{\mu}] \Big) S_{\mu}
           \tr \Big( h_+ B \Big) \bigg\} \no \\ 
& - & 
2 \, \ci \, \Big( \, h_{44} +  \mnod h_{60} \, \Big) \, 
     \bigg\{ \tr \Big( \bar{B} h_+ \Big)  S_{\mu}
\tr \Big( [ D^{\mu} , v \cdot u]  B \Big)
- \tr \Big(\bar{B} [ D^{\mu} , v \cdot u] \Big) S_{\mu}
           \tr \Big( h_+ B \Big) \bigg\} \no \\ 
& - & 
\ci \, \mnod h_{47} \, \bigg\{ \tr \Big( \bar{B}  
                     [ h_+,[ [ v \cdot D, v \cdot u], B] ] \Big) 
- \tr \Big( \bar{B} 
          [ [v \cdot D, v \cdot u ],[ h_+ , B] ] \Big) \bigg\} \no \\ 
& - & 
\ci \, \mnod \, h_{49} \bigg\{ \tr \Big( \bar{B}  
                     [ h_+,\{ [ v \cdot D, v \cdot u], B\} ] \Big) 
- \tr \Big( \bar{B} 
          \{ [v \cdot D, v \cdot u ],[ h_+ , B] \} \Big) \bigg\} \no \\ 
& - & 
\ci \, \mnod h_{52} \, \bigg\{ \tr \Big( \bar{B} h_+ \Big)  
\tr \Big( [v \cdot D, v \cdot u  ]  B \Big)
- \tr \Big(\bar{B} [v \cdot D, v \cdot u  ] \Big) 
           \tr \Big( h_+ B \Big) \bigg\}  
\eeqa
The relativistic corrections are
\beqa
& & {\cal L}_{\phi B}^{W \, (2,rc)}  \no \\ 
& = &
\frac{1}{\mnod} \, g_3 \, \bigg[
\frac{\ci}{2} \, \tr \Big( \bar{B} [ h_+, [[D_{\mu},u^{\mu}],B]]\Big)
+ \ci \, \tr \Big( \bar{B} [ h_+, [u^{\mu},[D_{\mu},B]]]\Big)\no \\
& + &
\frac{\ci}{2} \, \tr \Big( \bar{B} [ [D_{\mu},u^{\mu}], [h_+,B]]\Big)
+ \ci \, \tr \Big( \bar{B} [u^{\mu} , [h_+,[D_{\mu},B]]]\Big)\no \\
& - &
\frac{\ci}{2} \, \tr \Big( \bar{B} [ h_+, [[v \cdot D,v \cdot u],B]]\Big)
- \ci \, \tr \Big( \bar{B} [ h_+, [v \cdot u,[v \cdot D,B]]]\Big)\no \\
& - &
\frac{\ci}{2} \, \tr \Big( \bar{B} [ [v \cdot D,v \cdot u], [h_+,B]]\Big)
- \ci \, \tr \Big( \bar{B} [v \cdot u , [h_+,[v \cdot D,B]]]\Big)\no \\
& + &
\epsilon_{\mu \nu \alpha \beta} \, v^{\alpha}  \bigg\{
\tr \Big( \bar{B} S^{\beta} [ h_+ ,[[D^{\mu}, u^{\nu}],B]] \Big)
+ \tr \Big( \bar{B} S^{\beta} [ [D^{\mu}, u^{\nu}] ,[h_+,B]] \Big) \bigg\}
\bigg] \no \\
& + &
\frac{1}{\mnod} \, g_5 \, \bigg[
\frac{\ci}{2} \, \tr \Big( \bar{B} [ h_+, \{[D_{\mu},u^{\mu}],B\}]\Big)
+ \ci \, \tr \Big( \bar{B} [ h_+, \{u^{\mu},[D_{\mu},B]]\}]\Big)\no \\
& + &
\frac{\ci}{2} \, \tr \Big( \bar{B} \{ [D_{\mu},u^{\mu}], [h_+,B]\}\Big)
+ \ci \, \tr \Big( \bar{B} \{u^{\mu} , [h_+,[D_{\mu},B]]\}\Big)\no \\
& - &
\frac{\ci}{2} \, \tr \Big( \bar{B} [ h_+, \{[v \cdot D,v \cdot u],B\}]\Big)
- \ci \, \tr \Big( \bar{B} [ h_+, \{v \cdot u,[v \cdot D,B]\}]\Big)\no \\
& - &
\frac{\ci}{2} \, \tr \Big( \bar{B} \{ [v \cdot D,v \cdot u], [h_+,B]\}\Big)
- \ci \, \tr \Big( \bar{B} \{v \cdot u , [h_+,[v \cdot D,B]]\} \Big)\no \\
& + &
\epsilon_{\mu \nu \alpha \beta} \, v^{\alpha}  \bigg\{
\tr \Big( \bar{B} S^{\beta} [ h_+ ,\{[D^{\mu}, u^{\nu}],B\}] \Big)
+ \tr \Big( \bar{B} S^{\beta} \{ [D^{\mu}, u^{\nu}] ,[h_+,B]\} \Big) \bigg\}
\bigg] \no \\
& + &
\frac{1}{\mnod} \, g_7 \, \bigg[
\frac{\ci}{2} \, \tr \Big( \bar{B} \{ h_+, [[D_{\mu},u^{\mu}],B]\}\Big)
+ \ci \, \tr \Big( \bar{B} \{ h_+, [u^{\mu},[D_{\mu},B]]\}\Big)\no \\
& + &
\frac{\ci}{2} \, \tr \Big( \bar{B} [ [D_{\mu},u^{\mu}], \{h_+,B\}]\Big)
+ \ci \, \tr \Big( \bar{B} [u^{\mu} , \{h_+,[D_{\mu},B]\}]\Big)\no \\
& - &
\frac{\ci}{2} \, \tr \Big( \bar{B} \{ h_+, [[v \cdot D,v \cdot u],B]\}\Big)
- \ci \, \tr \Big( \bar{B} \{ h_+, [v \cdot u,[v \cdot D,B]]\}\Big)\no \\
& - &
\frac{\ci}{2} \, \tr \Big( \bar{B} [ [v \cdot D,v \cdot u], \{h_+,B\}]\Big)
- \ci \, \tr \Big( \bar{B} [v \cdot u , \{h_+,[v \cdot D,B]\}]\Big)\no \\
& + &
\epsilon_{\mu \nu \alpha \beta} \, v^{\alpha}  \bigg\{
\tr \Big( \bar{B} S^{\beta} \{ h_+ ,[[D^{\mu}, u^{\nu}],B]\} \Big)
+ \tr \Big( \bar{B} S^{\beta} [ [D^{\mu}, u^{\nu}] ,\{h_+,B\}] \Big) \bigg\}
\bigg] \no \\
& + &
\frac{1}{\mnod} \, g_8 \, \bigg[
\frac{\ci}{2} \, \tr \Big( \bar{B} h_+ \Big)\tr \Big( [D_{\mu},u^{\mu}] B \Big)
+ \ci \, \tr \Big( \bar{B} h_+ \Big)\tr \Big( u^{\mu}[D_{\mu},B]\Big) \no \\
& + &
\frac{\ci}{2} \, \tr \Big( \bar{B} [D_{\mu},u^{\mu}] \Big)\tr \Big( h_+ B \Big)
+ \ci \, \tr \Big( \bar{B} u^{\mu}\Big)\tr \Big( h_+ [D_{\mu},B]\Big) \no \\
& - &
\frac{\ci}{2} \, \tr \Big( \bar{B} h_+ \Big)\tr 
                  \Big( [v \cdot D,v \cdot u] B \Big)
- \ci \, \tr \Big( \bar{B} h_+ \Big)\tr \Big( v \cdot u [v \cdot D,B]\Big) \no \\
& - &
\frac{\ci}{2} \, \tr \Big( \bar{B} 
         [v \cdot D,v \cdot u] \Big)\tr \Big( h_+ B \Big)
- \ci \, \tr \Big( \bar{B} v \cdot u \Big)\tr \Big( h_+ [v \cdot D,B]\Big) \no \\
& + &
\epsilon_{\mu \nu \alpha \beta} \, v^{\alpha}  \bigg\{
\tr \Big( \bar{B} h_+ \Big) S^{\beta} \tr \Big( [D^{\mu}, u^{\nu}] B \Big)
+ \tr \Big( \bar{B} [D^{\mu}, u^{\nu}] \Big) S^{\beta} \tr \Big( h_+ B \Big)
\bigg] \no \\
& + &
\frac{1}{\mnod} \, g_{10} \, \bigg[
\frac{\ci}{2} \, \tr \Big( \bar{B} B \Big) \tr \Big( h_+ [D_{\mu},u^{\mu}]\Big)
+ \ci \, \tr \Big( \bar{B} [D_{\mu},B]\Big) \tr \Big( h_+ u^{\mu}\Big)\no \\
& - &
\frac{\ci}{2} \, \tr \Big( \bar{B} B \Big) \tr 
            \Big( h_+ [v \cdot D,v \cdot u]\Big)
- \ci\,  \tr \Big( \bar{B} [v \cdot D,B]\Big) \tr \Big( h_+ v \cdot u \Big)\no \\
& + &
\epsilon_{\mu \nu \alpha \beta} \, v^{\alpha}  \, \tr \Big( \bar{B} 
 S^{\beta} B \Big) \tr \Big( h_+ [D^{\mu},u^{\nu}]\Big)
\bigg] \no \\
& - &
\frac{\ci}{\mnod} \, g_{11} \, \bigg[
\tr \Big( \bar{B}  S_{\mu} [ h_+ , [[ D^{\mu}, v \cdot u ],B]]\Big)
+ 2 \tr \Big( \bar{B}  S_{\mu} [ h_+ , [ v \cdot u ,[ D^{\mu},B]]]\Big)\no \\
& + &
\tr \Big( \bar{B}  S_{\mu} [ [ D^{\mu}, v \cdot u ] , [h_+,B]]\Big)
+ 2 \tr \Big( \bar{B}  S_{\mu} [ v \cdot u  , [h_+ ,[ D^{\mu},B]]]\Big)
\bigg] \no \\
& - &
\frac{\ci}{\mnod} \, g_{13} \, \bigg[
\tr \Big( \bar{B}  S_{\mu} [ h_+ , \{[ D^{\mu}, v \cdot u ],B\}]\Big)
+ 2 \tr \Big( \bar{B}  S_{\mu} [ h_+ , \{ v \cdot u ,[ D^{\mu},B]\}]\Big)\no \\
& + &
\tr \Big( \bar{B}  S_{\mu} \{ [ D^{\mu}, v \cdot u ] , [h_+,B]\}\Big)
+ 2 \tr \Big( \bar{B}  S_{\mu} \{ v \cdot u  , [h_+ ,[ D^{\mu},B]]\}\Big)
\bigg] \no \\
& - &
\frac{\ci}{\mnod} \, g_{15} \, \bigg[
\tr \Big( \bar{B}  S_{\mu} \{ h_+ , [[ D^{\mu}, v \cdot u ],B]\}\Big)
+ 2 \tr \Big( \bar{B}  S_{\mu} \{ h_+ , [ v \cdot u ,[ D^{\mu},B]]\}\Big)\no \\
& + &
\tr \Big( \bar{B}  S_{\mu} [ [ D^{\mu}, v \cdot u ] , \{h_+,B\}]\Big)
+ 2 \tr \Big( \bar{B}  S_{\mu} [ v \cdot u  , \{h_+ ,[ D^{\mu},B]\}]\Big)
\bigg] \no \\
& - &
\frac{\ci}{\mnod} \, g_{16} \, \bigg[
\tr \Big( \bar{B} h_+ \Big)  S_{\mu}\tr \Big( [ D^{\mu},v \cdot u ]B \Big)
+ 2 \tr \Big( \bar{B} h_+ \Big)  S_{\mu}\tr \Big( v \cdot u [ D^{\mu},B] \Big)
\no \\
& + &
\tr \Big( \bar{B} [ D^{\mu},v \cdot u ] \Big)  S_{\mu}\tr \Big( h_+ B \Big)
+ 2 \tr \Big( \bar{B} v \cdot u \Big)  S_{\mu}\tr \Big( h_+ [ D^{\mu},B] \Big)
\bigg] \no \\
& - &
\frac{\ci}{\mnod} \, g_{18} \, \bigg[
\tr \Big( \bar{B}  S_{\mu} B \Big) \tr \Big( h_+ [ D^{\mu},v \cdot u ] \Big)
+ 2 \tr \Big( \bar{B} S_{\mu} [ D^{\mu},B] \Big) \tr \Big( h_+ v \cdot u \Big)
\bigg] \no \\
& - &
\frac{1}{4 \, \mnod^2} \, d \, \bigg[
\tr \Big( \bar{B} \{ [ D_{\mu},h_+], [ D^{\mu},B]\} \Big)
+ \tr \Big( \bar{B} \{ h_+, [ D_{\mu},[ D^{\mu},B]]\} \Big)\no \\
& - &
\tr \Big( \bar{B} \{ [ v \cdot D,h_+], [ v \cdot D,B]\} \Big)
- \tr \Big( \bar{B} \{ h_+, [ v \cdot D,[v \cdot  D,B]]\} \Big)\no \\
& - &
2 \ci \epsilon_{\mu \nu \alpha \beta} \, v^{\alpha}  
\tr \Big( \bar{B} S^{\beta} \{ [D^{\mu}, h_+] ,[D^{\nu},B]\} \Big)
\bigg] \no \\
& - &
\frac{1}{4 \, \mnod^2} \, f \, \bigg[
\tr \Big( \bar{B} [ [ D_{\mu},h_+], [ D^{\mu},B]] \Big)
+ \tr \Big( \bar{B} [ h_+, [ D_{\mu},[ D^{\mu},B]]] \Big)\no \\
& - &
\tr \Big( \bar{B} [ [ v \cdot D,h_+], [ v \cdot D,B]] \Big)
- \tr \Big( \bar{B} [ h_+, [ v \cdot D,[v \cdot  D,B]]] \Big)\no \\
& - &
2 \ci \epsilon_{\mu \nu \alpha \beta} \, v^{\alpha}  
\tr \Big( \bar{B} S^{\beta} [ [D^{\mu}, h_+] ,[D^{\nu},B]] \Big)
\bigg] \no \\
& - &
\frac{\ci}{4 \, \mnod^2} \, D \, d \, \bigg[
\tr \Big( \bar{B} S_{\mu} \{ h_+,\{  [ D^{\mu},v \cdot u ],B\}\} \Big)
+ \tr \Big( \bar{B} S_{\mu} \{ h_+,\{  v \cdot u ,[ D^{\mu},B]\}\} \Big)\no \\
& + &
\tr \Big( \bar{B} S_{\mu} \{  v \cdot u ,\{h_+ ,[ D^{\mu},B]\}\} \Big)
\bigg] \no \\
& + &
\frac{\ci}{3 \, \mnod^2} \, D \, d \, \bigg[
\tr \Big( \bar{B} h_+\Big) S_{\mu}\tr \Big([ D^{\mu},v \cdot u ],B \Big)
+ \tr \Big( \bar{B} h_+\Big) S_{\mu}  
        \tr \Big(v \cdot u ,[ D^{\mu},B] \Big)\no \\
& + &
\tr \Big( \bar{B} v \cdot u \Big) S_{\mu}\tr \Big(h_+ ,[ D^{\mu},B]\Big)
\bigg] \no \\
& - &
\frac{\ci}{4 \, \mnod^2} \, D \, f \, \bigg[
\tr \Big( \bar{B} S_{\mu} [ h_+,\{  [ D^{\mu},v \cdot u ],B\}] \Big)
+ \tr \Big( \bar{B} S_{\mu} [ h_+,\{  v \cdot u ,[ D^{\mu},B]\}] \Big)\no \\
& + &
\tr \Big( \bar{B} S_{\mu} \{  v \cdot u ,[h_+ ,[ D^{\mu},B]]\} \Big)
\bigg] \no \\
& - &
\frac{\ci}{4 \, \mnod^2} \, F \, d \, \bigg[
\tr \Big( \bar{B} S_{\mu} \{ h_+,[  [ D^{\mu},v \cdot u ],B]\} \Big)
+ \tr \Big( \bar{B} S_{\mu} \{ h_+,[  v \cdot u ,[ D^{\mu},B]]\} \Big)\no \\
& + &
\tr \Big( \bar{B} S_{\mu} [  v \cdot u ,\{h_+ ,[ D^{\mu},B]\}] \Big)
\bigg] \no \\
& - &
\frac{\ci}{4 \, \mnod^2} \, F \, f \, \bigg[
\tr \Big( \bar{B} S_{\mu} [ h_+,[  [ D^{\mu},v \cdot u ],B]] \Big)
+ \tr \Big( \bar{B} S_{\mu} [ h_+,[  v \cdot u ,[ D^{\mu},B]]] \Big)\no \\
& + &
\tr \Big( \bar{B} S_{\mu} [  v \cdot u ,[h_+ ,[ D^{\mu},B]]] \Big)
\bigg] 
\eeqa
We have not absorbed some of the relativistic corrections into 
$ {\cal L}_{\phi B}^{W \, (2,br)}  $ or the
$ O_i^{(2)} $.
\\
Finally, the strong meson-baryon Lagrangian reads
\beqa
{\cal L}_{\phi B} &=&
 {\cal L}_{\phi B}^{(1)} + {\cal L}_{\phi B}^{(2)} 
                  + {\cal L}_{\phi B}^{(3)} \\
&& \no \\
{\cal L}_{\phi B}^{(1)}  
& = &
 \ci \tr \Big( \bar{B} [ v \cdot D , B] \Big) +            
D \tr \Big( \bar{B} S_{\mu} \{ u^{\mu}, B\} \Big) 
+ F \tr \Big( \bar{B} S_{\mu} [ u^{\mu}, B] \Big) \\ 
&& \no \\
{\cal L}_{\phi B}^{(2)}  & = & {\cal L}_{\phi B}^{(2,rc)} \no \\
& = &
- \frac{1}{2 \mnod} \tr \Big( \bar{B} [ D_{\mu}, [D^{\mu}, B]] \Big) 
+ \frac{1}{2 \mnod} \tr \Big( \bar{B} [ v \cdot D, [v \cdot D, B]] \Big) \no \\
& - &
\frac{\ci}{2 \mnod} \, D \,  \tr \Big( \bar{B} S_{\mu} 
\{ [D^{\mu}, v \cdot u ] , B \} \Big)
- \frac{\ci}{\mnod} \, D \,  \tr \Big( \bar{B} S_{\mu} 
\{ v \cdot u  , [D^{\mu}, B ]\} \Big)\no \\
& - &
\frac{\ci}{2 \mnod} \, F \,  \tr \Big( \bar{B} S_{\mu} 
[ [D^{\mu}, v \cdot u ] , B ] \Big)
- \frac{\ci}{\mnod} \, F \,  \tr \Big( \bar{B} S_{\mu} 
[ v \cdot u  , [D^{\mu}, B ]] \Big) \no \\
&&
\eeqa
\beq
{\cal L}_{\phi B}^{(3)}    = 
{\cal L}_{\phi B}^{(3,br)} \: +  \:
\sum_{i} \, H_i O_i^{(3)} \: +  \: 
{\cal L}_{\phi B}^{(3,rc)} 
\eeq
\beqa \label{eq:b17}
{\cal L}_{\phi B}^{(3,br)}  
& = &
2 \, H_4 \bigg\{ \tr \Big( \bar{B} S_{\mu} 
                    [ \chi_+ , [u^{\mu}, B ]] \Big)
+ \tr \Big( \bar{B} S_{\mu} 
                    [ u^{\mu} , [\chi_+, B ]] \Big)\bigg\} \no \\
& + &
2 \, H_5 \bigg\{ \tr \Big( \bar{B} S_{\mu} 
                    [ \chi_+ , \{u^{\mu}, B \}] \Big)
+ \tr \Big( \bar{B} S_{\mu} 
                    \{ u^{\mu} , [\chi_+, B ]\} \Big)\bigg\} \no \\
& + &
2 \, H_6 \bigg\{ \tr \Big( \bar{B} S_{\mu} 
                    \{ \chi_+ , [u^{\mu}, B ]\} \Big)
+ \tr \Big( \bar{B} S_{\mu} 
                    [ u^{\mu} , \{ \chi_+, B \} ] \Big)\bigg\} \no \\
& + &
2 \, H_7 \bigg\{ \tr \Big( \bar{B} \chi_+ \Big)  S_{\mu} 
                  \tr \Big( u^{\mu} B \Big) 
+  \tr \Big( \bar{B} u^{\mu} \Big)  S_{\mu} 
                  \tr \Big( \chi_+ B \Big) \bigg\} \no \\
& + &
2 \, H_8 \tr \Big( \bar{B} S_{\mu}   [ u^{\mu}, B] \Big)
                \tr \Big( \chi_+ \Big) 
+
2 \, H_9 \tr \Big( \bar{B} S_{\mu}  \{ u^{\mu},B\} \Big)
                \tr \Big( \chi_+ \Big) \no \\
& + &
2 \, H_{10} \tr \Big( \bar{B} S_{\mu}  B \Big)
                \tr \Big( \chi_+ u^{\mu} \Big)\no \\
& + &
2 \, \ci \, H_{19} \, \tr \Big( \bar{B} S_{\mu}
[[ D^{\mu}, \chi_-], B]   \Big) 
+ 2 \, \ci \, H_{20} \, \tr \Big( \bar{B} S_{\mu}
\{[ D^{\mu}, \chi_-], B\}   \Big) \no \\ 
& + &
 2 \, \ci \, H_{21}\, \tr \Big( \bar{B} S_{\mu} B \Big)
\tr \Big( [ D^{\mu}, \chi_-]\Big) 
\no \\ 
& + &
\ci \, x_1 \, \tr \Big( \bar{B} \{ \chi_+ , [ v \cdot D , B ]\} \Big) 
+
\ci \, x_2 \, \tr \Big( \bar{B} [ \chi_+ , [ v \cdot D , B ]] \Big) \no \\
& + &
\ci \, x_3 \, 
\tr \Big( \bar{B} [ v \cdot D , B ] \Big) \tr \Big(\chi_+\Big)   
\eeqa
The last three terms renormalize the momentum dependent divergences
of the self--energy diagrams and, therefore, contribute to the $Z$-factors,
see app.~C.
\beqa
\sum_{i} \, H_i O_i^{(3)} & = &
2 H_{13} \, \tr   \Big( \bar{B} S_{\mu}  
[[ D^{\nu}, [D_{\nu},u^{\mu}]], B ] \Big) 
+ 2 H_{14} \, \tr   \Big( \bar{B} S_{\mu}  
\{[ D^{\nu}, [D_{\nu},u^{\mu}]], B \} \Big) \no \\
& + &
2 \mnod \Big( H_{15} - \mnod H_{17} \Big) 
\tr   \Big( \bar{B} S_{\mu}  [[v \cdot D, [v \cdot D,u^{\mu}]], B ] 
\Big) \no \\
& - &
2 \mnod H_{15} \tr   \Big( \bar{B} S_{\mu}  
[[D^{\mu}, [v \cdot D,v \cdot u]], B ] \Big) \no \\
& + &
2 \mnod \Big( H_{16} - \mnod H_{18} \Big) 
\tr   \Big( \bar{B} S_{\mu}  \{[v \cdot D, [v \cdot D,u^{\mu}]], B \} 
\Big) \no \\
& - &  2 \mnod H_{16} \tr   \Big( \bar{B} S_{\mu}  
\{[D^{\mu}, [v \cdot D,v \cdot u]], B \} \Big) 
\eeqa
\beqa
&& {\cal L}_{\phi B}^{(3,rc)} \no \\
& = &
\frac{\ci}{4 \mnod^2} \tr \Big( \bar{B} 
            [ D_{\mu}, [D^{\mu},[v \cdot D, B]]] \Big) 
+ \frac{\ci}{4 \mnod^2} \tr \Big( \bar{B} 
           [ v \cdot D, [v \cdot D,[v \cdot D, B]]] \Big) \no \\
& + &
\frac{\ci}{8 \mnod^2} \, D \, \epsilon_{\mu \nu \alpha \beta} v^{\beta}
\,  \tr \Big( \bar{B} \{ [D^{\mu}, u^{\alpha}], [ D^{\nu},B] \} \Big) \no \\
& + &
\frac{1}{8 \mnod^2} \, D \, \epsilon_{\nu \alpha \beta \gamma} 
\epsilon^{\mu \delta \rho \gamma}   v^{\beta} v_{\rho}
\,  \tr \Big( \bar{B} S_{\delta} \{ [D_{\mu}, u^{\alpha}], 
   [ D^{\nu},B] \} \Big) \no \\
& + &
\frac{1}{8 \mnod^2} \, D \, \epsilon_{\nu \alpha \beta \gamma} 
\epsilon^{\mu \delta \rho \gamma}   v^{\beta} v_{\rho}
\,  \tr \Big( \bar{B} S_{\delta} \{ u^{\alpha}, 
   [D_{\mu}, [ D^{\nu},B]] \} \Big) \no \\
& + &
\frac{1}{8 \mnod^2} \, D \, \epsilon_{\mu \alpha \beta \gamma} 
\epsilon^{\nu \delta \rho \gamma}   v^{\beta} v_{\rho}
\,  \tr \Big( \bar{B} S_{\delta} \{ [D^{\mu}, u^{\alpha}], 
   [ D_{\nu},B] \} \Big) \no \\
& + &
\frac{1}{8 \mnod^2} \, D \, \epsilon_{\mu \alpha \beta \gamma} 
\epsilon^{\nu \delta \rho \gamma}   v^{\beta} v_{\rho}
\,  \tr \Big( \bar{B} S_{\delta} \{ u^{\alpha}, 
   [D^{\mu}, [ D_{\nu},B]] \} \Big) \no \\
& + &
\frac{1}{8 \mnod^2} \, D \, \tr \Big( \bar{B} S_{\mu} [D^{\mu},\{ u_{\nu},
[D^{\nu},B] \} ] \Big) +
\frac{1}{8 \mnod^2} \, D \, \tr \Big( \bar{B} S_{\mu} [D^{\nu},\{ u_{\nu},
[D^{\mu},B] \} ] \Big) \no \\
& - &
\frac{1}{8 \mnod^2} \, D \, \tr \Big( \bar{B} S_{\mu} [D^{\mu},\{v \cdot u,
[v \cdot D,B] \} ] \Big) -
\frac{1}{8 \mnod^2} \, D \, \tr \Big( \bar{B} S_{\mu} [v \cdot D,\{ v \cdot u,
[D^{\mu},B] \} ] \Big) \no \\
& + &
\frac{\ci}{8 \mnod^2} \, F \, \epsilon_{\mu \nu \alpha \beta} v^{\beta}
\,  \tr \Big( \bar{B} [ [D^{\mu}, u^{\alpha}], [ D^{\nu},B] ] \Big) \no \\
& + &
\frac{1}{8 \mnod^2} \, F \, \epsilon_{\nu \alpha \beta \gamma} 
\epsilon^{\mu \delta \rho \gamma}   v^{\beta} v_{\rho}
\,  \tr \Big( \bar{B} S_{\delta} [ [D_{\mu}, u^{\alpha}], 
   [ D^{\nu},B] ] \Big) \no \\
& + &
\frac{1}{8 \mnod^2} \, F \, \epsilon_{\nu \alpha \beta \gamma} 
\epsilon^{\mu \delta \rho \gamma}   v^{\beta} v_{\rho}
\,  \tr \Big( \bar{B} S_{\delta} [ u^{\alpha}, 
   [D_{\mu}, [ D^{\nu},B]] ] \Big) \no \\
& + &
\frac{1}{8 \mnod^2} \, F\, \epsilon_{\mu \alpha \beta \gamma} 
\epsilon^{\nu \delta \rho \gamma}   v^{\beta} v_{\rho}
\,  \tr \Big( \bar{B} S_{\delta} [ [D^{\mu}, u^{\alpha}], 
   [ D_{\nu},B] ] \Big) \no \\
& + &
\frac{1}{8 \mnod^2} \, F \, \epsilon_{\mu \alpha \beta \gamma} 
\epsilon^{\nu \delta \rho \gamma}   v^{\beta} v_{\rho}
\,  \tr \Big( \bar{B} S_{\delta} [ u^{\alpha}, 
   [D^{\mu}, [ D_{\nu},B]] ] \Big) \no \\
& + &
\frac{1}{8 \mnod^2} \, F \, \tr \Big( \bar{B} S_{\mu} [D^{\mu},[ u_{\nu},
[D^{\nu},B] ] ] \Big) +
\frac{1}{8 \mnod^2} \, F \, \tr \Big( \bar{B} S_{\mu} [D^{\nu},[ u_{\nu},
[D^{\mu},B] ] ] \Big) \no \\
& - &
\frac{1}{8 \mnod^2} \, F \, \tr \Big( \bar{B} S_{\mu} [D^{\mu},[v \cdot u,
[v \cdot D,B] ] ] \Big) -
\frac{1}{8 \mnod^2} \, F \, \tr \Big( \bar{B} S_{\mu} [v \cdot D,[ v \cdot u,
[D^{\mu},B] ] ] \Big) \no \\
& - &
\frac{1}{4 \mnod^2} \, D \, \tr \Big( \bar{B} S_{\mu}
\{ [ D^{\mu},[v \cdot D,v \cdot u]], B \} \Big) -
\frac{1}{4 \mnod^2} \, D \, \tr \Big( \bar{B} S_{\mu}
\{ [v \cdot D,v \cdot u],[ D^{\mu}, B] \} \Big) \no \\ 
& - &
\frac{1}{4 \mnod^2} \, D \, \tr \Big( \bar{B} S_{\mu}
\{ [ D^{\mu},v \cdot u],[v \cdot D, B] \} \Big) -
\frac{1}{2 \mnod^2} \, D \, \tr \Big( \bar{B} S_{\mu}
\{ v \cdot u,[ D^{\mu},[v \cdot D, B]] \} \Big) \no \\ 
& - &
\frac{1}{4 \mnod^2} \, F \, \tr \Big( \bar{B} S_{\mu}
[ [ D^{\mu},[v \cdot D,v \cdot u]], B ] \Big) -
\frac{1}{4 \mnod^2} \, F \, \tr \Big( \bar{B} S_{\mu}
[ [v \cdot D,v \cdot u],[ D^{\mu}, B] ] \Big) \no \\ 
& - &
\frac{1}{4 \mnod^2} \, F \, \tr \Big( \bar{B} S_{\mu}
[ [ D^{\mu},v \cdot u],[v \cdot D, B] ] \Big) -
\frac{1}{2 \mnod^2} \, F \, \tr \Big( \bar{B} S_{\mu}
[ v \cdot u,[ D^{\mu},[v \cdot D, B]] ] \Big) \no \\ 
& + &
\frac{\ci}{\mnod} \, H_1 \, \tr \Big( \bar{B} S_{\mu}
[[ D^{\mu}, \chi_-], B]   \Big) 
+ \frac{\ci}{\mnod} \, H_2 \, \tr \Big( \bar{B} S_{\mu}
\{[ D^{\mu}, \chi_-], B\}   \Big) \no \\ 
& + &
\frac{\ci}{\mnod} \, H_3 \, \tr \Big( \bar{B} S_{\mu} B \Big)
\tr \Big( [ D^{\mu}, \chi_-]\Big) 
\no \\ 
& - &
\frac{1}{\mnod} \, H_{11} \,  
\tr \Big( \bar{B} S_{\mu} [[ D^{\mu}, [D^{\nu},u_{\nu}]], B ]\Big)
- \frac{1}{\mnod} \, H_{12} \,  
\tr \Big( \bar{B} S_{\mu} \{[ D^{\mu}, [D^{\nu},u_{\nu}]], B \}\Big)
\eeqa

\def\theequation{\Alph{section}.\arabic{equation}}
\setcounter{equation}{0}
\section{Z-factors} \label{app.c}
In this appendix we display explicit expressions for the $Z$-factors
and the chiral correction at next order to the pseudoscalar decay constant.
\beqa
Z_\Sigma  & = &
1 - \frac{1}{\Lambda_\chi^2} \, \bigg\{
\Big( \frac{2}{3} D^2 + 4 F^2 \Big) M_\pi^2 \Big( \frac{3}{2} 
\ln \left( \frac{ M_\pi^2}{\lambda^2} \right) +1 \Big)\no \\
& + &
2 \Big( D^2 + F^2 \Big) M_K^2 \Big( \frac{3}{2} 
\ln \left( \frac{ M_K^2}{\lambda^2} \right) +1 \Big) 
- \frac{2}{3} D^2  M_\eta^2  \Big( \frac{3}{2} 
\ln \left( \frac{ M_\eta^2}{\lambda^2} \right) +1 \Big) \bigg\}\no \\
& - &
x_1^r 4 M_\pi^2
- x_3^r \Big( 2 M_\pi^2 + 4 M_K^2\Big) \no \\
&& \no \\
Z_N  & = &
1 - \frac{1}{\Lambda_\chi^2} \, \bigg\{
\frac{3}{2} \Big( D + F  \Big)^2  M_\pi^2 \Big( \frac{3}{2} 
\ln \left( \frac{ M_\pi^2}{\lambda^2} \right) +1 \Big)\no \\
& + &
\Big[ \frac{3}{2} \Big( D - F  \Big)^2 
+ \frac{1}{6} \Big( D + 3F  \Big)^2 \Big] M_K^2 \Big( \frac{3}{2} 
\ln \left( \frac{ M_K^2}{\lambda^2} \right) +1 \Big) \no \\
& + &
\frac{1}{6} \Big( D - 3F  \Big)^2 M_\eta^2  \Big( \frac{3}{2} 
\ln \left( \frac{ M_\eta^2}{\lambda^2} \right) +1 \Big) \bigg\}
\,+ \,  \frac{1}{4 m_N^2} M_\pi^2 \no \\
& - &
x_1^r 4 M_K^2
- x_2^r 4 \Big( M_\pi^2 - M_K^2\Big)
- x_3^r \Big( 2 M_\pi^2 + 4 M_K^2\Big) \no \\
&& \no \\
Z_\Xi  & = &
1 - \frac{1}{\Lambda_\chi^2} \, \bigg\{
\frac{3}{2} \Big( D - F  \Big)^2  M_\pi^2 \Big( \frac{3}{2} 
\ln \left( \frac{ M_\pi^2}{\lambda^2} \right) +1 \Big)\no \\
& + &
\Big[ \frac{3}{2} \Big( D + F  \Big)^2 
+ \frac{1}{6} \Big( D - 3F  \Big)^2 \Big] M_K^2 \Big( \frac{3}{2} 
\ln \left( \frac{ M_K^2}{\lambda^2} \right) +1 \Big) \no \\
& + &
\frac{1}{6} \Big( D + 3F  \Big)^2 M_\eta^2  \Big( \frac{3}{2} 
\ln \left( \frac{ M_\eta^2}{\lambda^2} \right) +1 \Big) \bigg\}\no \\
& - &
x_1^r 4 M_K^2
+ x_2^r 4 \Big( M_\pi^2 - M_K^2\Big)
- x_3^r \Big( 2 M_\pi^2 + 4 M_K^2\Big) \no \\
&& \no \\
Z_\Lambda  & = &
1 - \frac{1}{\Lambda_\chi^2} \, \bigg\{
2 D^2  M_\pi^2 \Big( \frac{3}{2} 
\ln \left( \frac{ M_\pi^2}{\lambda^2} \right) +1 \Big)\no \\
& + &
\frac{2}{3} \Big( D^2 + 9F^2  \Big) M_K^2 \Big( \frac{3}{2} 
\ln \left( \frac{ M_K^2}{\lambda^2} \right) +1 \Big) 
+ \frac{2}{3} D^2 M_\eta^2  \Big( \frac{3}{2} 
\ln \left( \frac{ M_\eta^2}{\lambda^2} \right) +1 \Big) \bigg\}\no \\
& + &
\frac{1}{4 m_\Lambda^2} M_\pi^2
\,-\, \frac{4}{3} x_1^r \Big( -M_\pi^2 + 4 M_K^2\Big) 
- x_3^r \Big( 2 M_\pi^2 + 4 M_K^2\Big) \no \\
&& \no \\
Z_\pi  & = &
1 + \frac{1}{\Lambda_\chi^2} \, \bigg\{
\frac{2}{3} M_\pi^2 \ln \left( \frac{ M_\pi^2}{\lambda^2} \right)
+ \frac{1 }{3} M_K^2 \ln \left( \frac{ M_K^2}{\lambda^2} \right)\bigg\}  \no \\
& + &
\frac{8}{F_{\pi}^2}\bigg\{ L_4^r \Big( 2 M_K^2 + M_\pi^2 \Big) 
+ L_5^r M_\pi^2 \bigg\}
\eeqa
where the $ L_i^r $ have been defined in \cite{GL1}, 
$\Lambda_\chi = 4 \pi F_{\pi}$
and the $ x_i^r $ represent the finite remainders of the LECs of
the following Lagrangian after renormalizing the momentum dependent
divergences of the sel--energy diagrams
\beqa
{\cal L} & = &
\ci \Big[ x_1^r + \frac{3 L}{4 F_\pi^2}\big( D^2 - 3 F^2 \big) \Big]
\tr \Big( \bar{B} \{ \chi_+ , [ v \cdot D , B ]\} \Big) \no \\
& + &
\ci \Big[ x_2^r - \frac{5 L}{2 F_\pi^2}\, DF \Big]
\tr \Big( \bar{B} [ \chi_+ , [ v \cdot D , B ]] \Big) \no \\
& + &
\ci \Big[ x_3^r - \frac{3 L}{2 F_\pi^2}\big( 
\frac{13}{9} D^2 +F^2 \big) \Big]
\tr \Big( \bar{B} [ v \cdot D , B ] \Big) \tr \Big(\chi_+\Big)   
\eeqa
Here, we set $x_i^r = 0 $.
Furthermore, one has to account for the contributions of the heavy
components of the external baryons to their $Z$-factors, see \cite{EM}.
In the rest frame of the heavy baryon they vanish for the
decaying baryon. For the light baryon with the mass $m_B$ we get a
term which is to lowest order $M_\pi^2 / (4 m_B^2)$. This factor has been
added to $Z_N$ and $Z_\Lambda$. In the case of $Z_\Lambda$ it has to be
neglected for the decay $\Lambda \rightarrow p \pi^-$.
\\
Finally, $ \delta F_\pi $ is defined via
\beq
F_\pi = \Fnod ( 1 +  \delta F_\pi )
\eeq
with
\beqa
\delta F_\pi & = &
- \frac{1}{\Lambda_\chi^2} \, \bigg\{
M_\pi^2 \ln \left( \frac{ M_\pi^2}{\lambda^2} \right) 
+ \frac{1}{2}\, M_K^2 \ln \left( \frac{ M_K^2}{\lambda^2} \right) 
\bigg\} \no \\
& + &
\frac{4}{F_{\pi}^2}\bigg\{ L_4^r \Big( 2 M_K^2 + M_\pi^2 \Big) 
+ L_5^r M_\pi^2 \bigg\} 
\eeqa
For $L_4^r$ and $L_5^r$ we use the central values of Bijnens et al. in 
ref.~\cite{daphne}.

\def\theequation{\Alph{section}.\arabic{equation}}
\setcounter{section}{3}
\setcounter{equation}{0}
\section{P-wave amplitudes} \label{app:d}
In this appendix we present the expressions for the coefficients of
the p-wave amplitudes.
\beqa
\alpha_{\Sigma^+ n}^{(p)} & = &  
- \frac{1}{m_\Sigma - m_N} 2 \, (D+F) \, (d-f)
+ \frac{1}{m_\Sigma - E_N} 2 \, F \, (d-f) -
 \frac{1}{m_\Lambda - E_N} \frac{2}{3} \, D \, (d+3f) \no \\
\beta_{\Sigma^+ n}^{(p)\,\pi} & = & 
- \frac{1}{m_\Sigma - m_N} 8 \Big( -h_3 - h_5 + h_7 \Big) (D+F)\no \\
& + &
\frac{1}{m_\Lambda - E_N} 8 \Big( -h_3 + \frac{1}{3} h_5 
                   - \frac{1}{3} h_7 + \frac{2}{3} h_8 \Big) \, D +
\frac{1}{m_\Sigma - E_N} 8 \Big( -h_3 - h_5 + h_7 \Big) \, F\no \\
& + &
\frac{1}{\Lambda_\chi^2} \, \frac{1}{m_\Sigma - m_N} 
\, 2 (D+F)^2 \,\Big( \, d (\frac{4}{3} D - F) + f \, F \Big) \no \\
& + &
\frac{1}{\Lambda_\chi^2} \, \frac{1}{m_\Lambda - E_N} 
\,\Big( \, d \big[ \frac{34}{9} D^3 + 2 D^2 F - \frac{8}{3} D F^2 \big]
+ \, f \, \big[ \frac{10}{3} D^3 + 2 D^2 F - 8 D F^2 \big] \Big)\no \\
& + &
\frac{1}{\Lambda_\chi^2} \, \frac{1}{m_\Sigma - E_N} 
\, \Big( \, d \big[ \frac{2}{3} D^2 F +\frac{10}{3} D F^2 - 8 F^3 \big]
+ \, f \, \big[-\frac{10}{3} D^2 F - 6 D F^2 + 8 F^3\big]\Big)\no \\
\beta_{\Sigma^+ n}^{(p)\,K} & = & 
\frac{1}{m_\Sigma - m_N} 8 \Big( -h_3 + h_5 + h_7 \Big) (D+F)\no \\
& - &
\frac{1}{m_\Lambda - E_N} 8 \Big( -h_3 + \frac{7}{3} h_5 
                   - \frac{1}{3} h_7 + \frac{2}{3} h_8 \Big) \, D -
\frac{1}{m_\Sigma - E_N} 8 \Big( -h_3 + h_5 + h_7 \Big) \, F \no \\
& + &
\frac{1}{\Lambda_\chi^2} \, \frac{1}{m_\Sigma - m_N} 
\,\Big( \, d \big[ \frac{8}{3} D^3 - \frac{16}{3} D F^2 + 8 F^3 \big]
+ \, f \, \big[ \frac{16}{3} D^2 F + 8 D F^2 - 8 F^3 \big] \Big)\no \\
& + &
\frac{1}{\Lambda_\chi^2} \, \frac{1}{m_\Lambda - E_N} 
\,\Big( \, d \big[ -\frac{4}{3} D^3 + 4 D^2 F + \frac{8}{3} D F^2 \big]
+ \, f \, \big[ - \frac{4}{3} D^3 + 4 D^2 F + 8 D F^2 \big] \Big)\no \\
& + &
\frac{1}{\Lambda_\chi^2} \, \frac{1}{m_\Sigma - E_N} 
\, \Big( \, d \big[ -\frac{8}{3} D^2 F +\frac{4}{3} D F^2 - 4 F^3 \big]
+ \, f \, \big[ - 4 D F^2 + 4 F^3\big]\Big)\no \\
\beta_{\Sigma^+ n}^{(p)\,\eta} & = & 
\frac{1}{\Lambda_\chi^2} \, \frac{1}{m_\Sigma - m_N} 
\,\Big( \, d \big[ \frac{4}{3} D^3 - \frac{14}{3} D^2 F + 6 F^3 \big]
+ \, f \, \big[- \frac{4}{3} D^3 +\frac{14}{3} D^2 F -6 F^3  \big] \Big)\no \\
& + &
\frac{1}{\Lambda_\chi^2} \, \frac{1}{m_\Lambda - E_N} 
\,\Big( \, d \big[ -\frac{2}{9} D^3 + \frac{2}{3}  D^2 F \big]
+ \, f \, \big[ - \frac{2}{3} D^3 + 2 D^2 F \big] \Big)\no \\
& + &
\frac{1}{\Lambda_\chi^2} \, \frac{1}{m_\Sigma - E_N} 
\, \Big( \, d \big[ -\frac{10}{3} D^2 F + 2 D F^2 \big]
+ \, f \, \big[ \frac{10}{3} D^2 F - 2 D F^2\big]\Big)\no \\
\gamma_{\Sigma^+ n}^{(p)\,\pi} & = & 
\frac{1}{m_\Sigma - m_N} \frac{17}{12} (d-f)(D+F)
+ \frac{1}{m_\Lambda - E_N} \frac{17}{36}  D \, (d+3f) -
\frac{1}{m_\Sigma - E_N} \frac{17}{12}  F \, (d-f)
\no \\ 
& + &
\frac{1}{m_\Sigma - m_N} 
\, (D+F)^2 \,\Big( \, d (5 D - 2F) + f \, ( - D +2F) \Big) \no \\
& + &
\frac{1}{m_\Lambda - E_N} 
\,\Big( \, d \big[ \frac{47}{9} D^3 + 3 D^2 F - \frac{4}{3} D F^2\big]
+ \, f \, \big[ \frac{11}{3} D^3 - 3 D^2 F - 4 D F^2\big] \Big)\no \\
& + &
\frac{1}{m_\Sigma - E_N} 
\, \Big( \, d \big[ -\frac{5}{3} D^2 F + 5 D F^2 - 8 F^3\big]
+ \, f \, \big[ -\frac{7}{3} D^2 F - 9 D F^2 + 8 F^3\big]\Big)\no \\
\gamma_{\Sigma^+ n}^{(p)\,K} & = & 
\frac{1}{m_\Sigma - m_N} \frac{11}{6} (d-f)(D+F)
+ \frac{1}{m_\Lambda - E_N} \frac{11}{18}  D \, (d+3f) -
\frac{1}{m_\Sigma - E_N} \frac{11}{6}  F \, (d-f)
\no \\ 
& + &
\frac{1}{m_\Sigma - m_N} 
\,\Big(\, d \, \big[ \frac{16}{3} D^3 -\frac{4}{3} D^2 F - 4 D F^2 
+8 F^3\big]\Big) \no \\ 
& + &
 f \, \big[ -\frac{4}{3} D^3 +\frac{28}{3} D^2 F +8 D F^2 
- 8 F^3\big]\Big) \no \\
& + &
\frac{1}{m_\Lambda - E_N} 
\,\Big(\, d \, \big[ -\frac{4}{3} D^3 + 6 D^2 F + \frac{10}{3} D F^2 
\big]\Big) 
+ f \, \big[ 6 D^2 F + \frac{10}{2} D F^2 \big]\Big) \no \\
& + &
\frac{1}{m_\Sigma - E_N} 
\, \Big( \, d \big[ -6 D^2 F + 2 D F^2 -4  F^3\big]
+ \, f \, \big[ 2 D^2 F - 6 D F^2 + 4 F^3\big]\Big)\no \\
\gamma_{\Sigma^+ n}^{(p)\,\eta} & = & 
\frac{1}{m_\Sigma - m_N} \frac{3}{4} (d-f)(D+F)
+ \frac{1}{m_\Lambda - E_N} \frac{1}{4}  D \, (d+3f) -
\frac{1}{m_\Sigma - E_N} \frac{3}{4}  F \, (d-f)
\no \\ 
& + &
\frac{1}{m_\Sigma - m_N} 
\,(d-f) \, (D+F) \,\Big(\, \frac{5}{3} D^2 - 7 DF + 6 F^2 \Big) \no \\
& + &
\frac{1}{m_\Lambda - E_N} ( d + 3f) \Big( \,\frac{1}{9} D^3 + D^2 F \Big)+
\frac{1}{m_\Sigma - E_N} (d-f)  \Big( \,-\frac{11}{3}D^2 F  + 3 DF^2\Big)\no \\
\epsilon_{\Sigma^+ n}^{(p)} & = &  
-8 g_{11} - 8 g_{13} + 8 g_{15} + 4  g_{16} \no \\
\delta_{\Sigma^+ n}^{(p)} & = &  
- \frac{1}{m_\Sigma - m_N} (D+F) \, (d-f)
- \frac{1}{m_\Lambda - E_N} \frac{1}{3}  D \, (d+3f) \no \\ 
& + &
\frac{1}{m_\Sigma - E_N}  F \, (d-f)\no \\ 
\rho_{\Sigma^+ n}^{(p)} & = &
- \frac{q^2 - (E_N - \mnod)^2}{(m_\Lambda - E_N)^2} \frac{1}{3}  D \, (d+3f)   
+ \frac{q^2 - (E_N - \mnod)^2}{(m_\Sigma - E_N)^2}  F \, (d-f)
\eeqa

The momentum of the outgoing baryon squared $ q^2 $ can be expressed
in terms of the physical masses. However, we retain the
notation $ q^2 $ for simplicity. \\
The coefficients for the other three decays read

\beqa
\alpha_{\Sigma^- n}^{(p)} & = &  
- \frac{1}{m_\Sigma - E_N} 2 \, F \, (d-f) 
- \frac{1}{m_\Lambda - E_N} \frac{2}{3} \, D \, (d+3f) \no \\
\beta_{\Sigma^- n}^{(p)\,\pi} & = & 
\frac{1}{m_\Lambda - E_N} 8 \Big( -h_3 + \frac{1}{3} h_5 
                   - \frac{1}{3} h_7 + \frac{2}{3} h_8 \Big) \, D +
\frac{1}{m_\Sigma - E_N} 8 \Big( h_3 + h_5 - h_7 \Big) \, F\no \\
& + &
\frac{1}{\Lambda_\chi^2} \, \frac{1}{m_\Lambda - E_N} 
\,\Big( \, d \big[ \frac{34}{9} D^3 + 2 D^2 F - \frac{8}{3} D F^2 \big]
+ \, f \, \big[ \frac{10}{3} D^3 + 2 D^2 F - 8 D F^2 \big] \Big)\no \\
& + &
\frac{1}{\Lambda_\chi^2} \, \frac{1}{m_\Sigma - E_N} 
\, \Big( \, d \big[- \frac{2}{3} D^2 F -\frac{10}{3} D F^2 + 8 F^3 \big]
+ \, f \, \big[\frac{10}{3} D^2 F + 6 D F^2 - 8 F^3\big]\Big)\no \\
\beta_{\Sigma^- n}^{(p)\,K} & = & 
- \frac{1}{m_\Lambda - E_N} 8 \Big( -h_3 + \frac{7}{3} h_5 
                   - \frac{1}{3} h_7 + \frac{2}{3} h_8 \Big) \, D -
\frac{1}{m_\Sigma - E_N} 8 \Big( h_3 - h_5 - h_7 \Big) \, F \no \\
& + &
\frac{1}{\Lambda_\chi^2} \, \frac{1}{m_\Lambda - E_N} 
\,\Big( \, d \big[ -\frac{4}{3} D^3 + 4 D^2 F + \frac{8}{3} D F^2 \big]
+ \, f \, \big[ - \frac{4}{3} D^3 + 4 D^2 F + 8 D F^2 \big] \Big)\no \\
& - &
\frac{1}{\Lambda_\chi^2} \, \frac{1}{m_\Sigma - E_N} 
\, \Big( \, d \big[ -\frac{8}{3} D^2 F +\frac{4}{3} D F^2 - 4 F^3 \big]
+ \, f \, \big[ - 4 D F^2 + 4 F^3\big]\Big)\no \\
\beta_{\Sigma^- n}^{(p)\,\eta} & = & 
\frac{1}{\Lambda_\chi^2} \, \frac{1}{m_\Lambda - E_N} 
\,\Big( \, d \big[ -\frac{2}{9} D^3 + \frac{2}{3}  D^2 F \big]
+ \, f \, \big[ - \frac{2}{3} D^3 + 2 D^2 F \big] \Big)\no \\
& + &
\frac{1}{\Lambda_\chi^2} \, \frac{1}{m_\Sigma - E_N} 
\, \Big( \, d \big[ \frac{10}{3} D^2 F - 2 D F^2 \big]
+ \, f \, \big[- \frac{10}{3} D^2 F + 2 D F^2\big]\Big)\no \\
\gamma_{\Sigma^- n}^{(p)\,\pi} & = & 
\frac{1}{m_\Lambda - E_N} \frac{17}{36}  D \, (d+3f) 
+ \frac{1}{m_\Sigma - E_N} \frac{17}{12}  F \, (d-f)
\no \\ 
& + &
\frac{1}{m_\Lambda - E_N} 
\,\Big( \, d \big[ \frac{47}{9} D^3 + 3 D^2 F - \frac{4}{3} D F^2\big]
+ \, f \, \big[ \frac{11}{3} D^3 - 3 D^2 F - 4 D F^2\big] \Big)\no \\
& + &
\frac{1}{m_\Sigma - E_N} 
\, \Big( \, d \big[ \frac{5}{3} D^2 F - 5 D F^2 + 8  F^3\big]
+ \, f \, \big[ \frac{7}{3} D^2 F + 9 D F^2 - 8  F^3\big]\Big)\no \\
\gamma_{\Sigma^- n}^{(p)\,K} & = & 
\frac{1}{m_\Lambda - E_N} \frac{11}{18}  D \, (d+3f)
+ \frac{1}{m_\Sigma - E_N} \frac{11}{6}  F \, (d-f)
\no \\ 
& + &
\frac{1}{m_\Lambda - E_N} 
\,\Big(\, d \, \big[ -\frac{4}{3} D^3 + 6 D^2 F + \frac{10}{3} D F^2 
\big]\Big) + f \, \big[  6 D^2 F + 10 D F^2 \big]\Big) \no \\
& + &
\frac{1}{m_\Sigma - E_N} 
\, \Big( \, d \big[ +6 D^2 F - 2 D F^2 +4  F^3\big]
+ \, f \, \big[ -2 D^2 F + 6 D F^2 - 4 F^3\big]\Big)\no \\
\gamma_{\Sigma^- n}^{(p)\,\eta} & = & 
\frac{1}{m_\Lambda - E_N} \frac{1}{4}  D \, (d+3f) 
+ \frac{1}{m_\Sigma - E_N} \frac{3}{4}  F \, (d-f)
\no \\ 
& + &
\frac{1}{m_\Lambda - E_N} ( d + 3f) \Big( \,\frac{1}{9} D^3 + D^2 F \Big)+
\frac{1}{m_\Sigma - E_N} (d-f)  \Big( \,\frac{11}{3}D^2 F  - 3 DF^2\Big)\no \\
\epsilon_{\Sigma^- n}^{(p)} & = &  
4 g_{11} - 4 g_{13} + 4 g_{15} \no \\
\delta_{\Sigma^- n}^{(p)} & = &  
- \frac{1}{m_\Lambda - E_N} \frac{1}{3}  D \, (d+3f) 
- \frac{1}{m_\Sigma - E_N}  F \, (d-f)\no \\ 
\rho_{\Sigma^- n}^{(p)} & = &
- \frac{q^2 - (E_N - \mnod)^2}{(m_\Lambda - E_N)^2} \frac{1}{3}  D \, (d+3f)   
- \frac{q^2 - (E_N - \mnod)^2}{(m_\Sigma - E_N)^2}  F \, (d-f)\no \\ 
\phi_{\Sigma^- n}^{(p)} & = &  D - F
\eeqa
\vskip 1.5cm
\beqa
\alpha_{\Lambda p}^{(p)} & = &  
\frac{1}{m_\Lambda-m_N} \frac{2}{\sqrt{6}} (d+3f)(D+F) \,
+ \frac{1}{m_\Sigma - E_N} \frac{4}{\sqrt{6}}\, D\,(d-f) \no \\
\beta_{\Lambda p}^{(p)\,\pi} & = & 
- \frac{1}{m_\Lambda - m_N} \frac{16}{\sqrt{6}} 
\Big( - \frac{3}{2} h_3 + \frac{1}{2} h_5 - \frac{1}{2} h_7 + h_8 \Big) \,
(D+F)\no \\
& + &
\frac{1}{m_\Sigma - E_N}\frac{16}{\sqrt{6}}
                \Big( -h_3 - h_5 + h_7 \Big) \, D\no \\
& + &
\frac{1}{\Lambda_\chi^2} \, \frac{1}{m_\Lambda - m_N} 
\, \frac{2}{\sqrt{6}}(D+F)^2 \,\Big( \, d (-4D - F) + 6f \,F\Big) \no \\
& + &
\frac{1}{\Lambda_\chi^2} \, \frac{1}{m_\Sigma - E_N} \frac{1}{\sqrt{6}}
\, \Big( \, d \big[ -\frac{20}{3} D^3 +\frac{20}{3} D^2 F +8 DF^2 \big]
+ \, f \, \big[ \frac{4}{3} D^3 - 12 D^2 F - 8 DF^2 \big]\Big)\no \\
\beta_{\Lambda p}^{(p)\,K} & = & 
\frac{1}{m_\Lambda - m_N} \frac{16}{\sqrt{6}} 
  \Big( -\frac{3}{2}h_3 + \frac{7}{2} h_5
              - \frac{1}{2} h_7 + h_8 \Big) (D+F)\no \\
& - &
\frac{1}{m_\Sigma - E_N}  \frac{16}{\sqrt{6}}
                  \Big( -h_3 + h_5 + h_7 \Big) \, D \no \\
& + &
\frac{1}{\Lambda_\chi^2} \, \frac{1}{m_\Lambda - m_N} \frac{2}{\sqrt{6}}
\,\Big( \, d \big[ \frac{2}{3} D^3 - \frac{14}{3}D^2 F -2DF^2 -2 F^3 \big]
\no \\
&& \; + \, f \, \big[ -2 D^3-6 D^2 F + 6 D F^2 - 6 F^3 \big] \Big)\no \\
& + &
\frac{1}{\Lambda_\chi^2} \, \frac{1}{m_\Sigma - E_N} \frac{2}{\sqrt{6}}
\, \Big( \, d \big[ -\frac{8}{3} D^3 +\frac{4}{3} D^2 F - 4 DF^2 \big]
+ \, f \, \big[ - 4 D^2 F + 4 D F^2\big]\Big)\no \\
\beta_{\Lambda p}^{(p)\,\eta} & = & 
\frac{1}{\Lambda_\chi^2} \, \frac{1}{m_\Lambda - m_N} \frac{2}{\sqrt{6}}
\,(d+3f) (D+F)(D-3F)F \no \\
& - &
\frac{1}{\Lambda_\chi^2} \, \frac{1}{m_\Sigma - E_N} \frac{4}{3\sqrt{6}}
\, (d-f) D^2 (D-3F) \no \\
\gamma_{\Lambda p}^{(p)\,\pi} & = & 
- \frac{1}{m_\Lambda - m_N}\frac{1}{\sqrt{6}} \frac{17}{12}  
                   \, (d+3f) (D+F)  -
\frac{1}{m_\Sigma - E_N} \frac{1}{\sqrt{6}}\frac{17}{6}  D \, (d-f)
\no \\ 
& - &
\frac{1}{m_\Lambda - m_N} \frac{1}{\sqrt{6}}
\, (D+F)^2 \,\Big( \, d (13D +4 F) + f \, ( 3 D + 12F) \Big) \no \\
& + &
\frac{1}{m_\Sigma - E_N} \frac{1}{\sqrt{6}}
\, \Big( \, d \big[ - \frac{22}{3} D^3 + 10 D^2 F - 4 DF^2\big]
+ \, f \, \big[ -\frac{2}{3} D^3 - 18 D^2 F +4 DF^2\big]\Big)\no \\
\gamma_{\Lambda p}^{(p)\,K} & = & 
- \frac{1}{m_\Lambda - m_N} \frac{1}{\sqrt{6}}\frac{11}{6}  
                         (D+F) \, (d+3f) -
\frac{1}{m_\Sigma - E_N}\frac{1}{\sqrt{6}} \frac{11}{3}  D \, (d-f)
\no \\ 
& - &
\frac{1}{m_\Lambda - m_N} 
\,\Big(\, d \, \big[- \frac{2}{3} D^3 +\frac{38}{3} D^2 F +10 D F^2 
+2 F^3\big]\Big) \no \\
&  &
+ f \, \big[ 10 D^3 -14 D^2 F -6 D F^2 +6 F^3\big]\Big) \no \\
& + &
\frac{1}{m_\Sigma - E_N} \frac{1}{\sqrt{6}}
\,\Big(\, d \, \big[ -12 D^3 + 4 D^2 F - 8 D F^2 \big] 
+ f \, \big[ 4 D^3 - 12 D^2 F + 8 D F^2\big]\Big) \no \\
\gamma_{\Lambda p}^{(p)\,\eta} & = & 
- \frac{1}{m_\Lambda - E_N} \frac{1}{\sqrt{6}}\frac{3}{4}(D+F)\, (d+3f) -
\frac{1}{m_\Sigma - E_N} \frac{1}{\sqrt{6}}\frac{3}{2}  D \, (d-f)
\no \\ 
& + &
\frac{1}{m_\Lambda - m_N} \frac{1}{\sqrt{6}}
      ( d + 3f) (D-3F) (D+F)(\frac{1}{3}D +2F)\no \\
& - &
\frac{1}{m_\Sigma - E_N}\frac{1}{\sqrt{6}} (d-f)  2 D^2 
           \Big( \,\frac{7}{3}D  - 3 F\Big)\no \\
\epsilon_{\Lambda p}^{(p)} & = &  
\frac{1}{\sqrt{6}} \Big( -12 g_{11} - 20 g_{13} - 4 g_{15} \Big) \no \\
\delta_{\Lambda p}^{(p)} & = &  
\frac{1}{m_\Lambda - E_N} \frac{1}{\sqrt{6}}  (D+F) \, (d+3f) +
\frac{1}{m_\Sigma - E_N}  2D \, (d-f)\no \\ 
\rho_{\Lambda p}^{(p)} & = &
\frac{1}{\sqrt{6}}
\frac{q^2 - (E_N - \mnod)^2}{(m_\Sigma - E_N)^2}  2D \, (d-f)\no \\ 
\phi_{\Lambda p}^{(p)} & = & - \frac{1}{\sqrt{6}} \, \Big( D+3F \Big)
\eeqa
\vskip 1.5cm
\beqa
\alpha_{\Xi^- \Lambda}^{(p)} & = &  
-\frac{1}{m_\Xi-E_\Lambda} \frac{2}{\sqrt{6}} (d-3f)(D-F) \,
- \frac{1}{m_\Xi-m_\Sigma} \frac{4}{\sqrt{6}}\, D\,(d+f) \no \\
\beta_{\Xi^- \Lambda}^{(p)\,\pi} & = & 
\frac{1}{m_\Xi-E_\Lambda} \frac{16}{\sqrt{6}} 
\Big( - \frac{3}{2} h_3 - \frac{1}{2} h_5 + \frac{1}{2} h_7 + h_8 \Big) \,
(D-F)\no \\
& - &
\frac{1}{m_\Xi-m_\Sigma}\frac{16}{\sqrt{6}}
                \Big( -h_3 + h_5 - h_7 \Big) \, D\no \\
& + &
\frac{1}{\Lambda_\chi^2} \, \frac{1}{m_\Xi-E_\Lambda} 
\, \frac{1}{\sqrt{6}} \,
 \Big( \, d \big[ 8 D^3 -18 D^2 F +12 DF^2 - 2 F^3\big]
+ \, f \, \big[ 6 D^2 F - 12 DF^2 + 6 F^3 \big]\Big) \no \\
& + &
\frac{1}{\Lambda_\chi^2} \, \frac{1}{m_\Xi-m_\Sigma} \frac{1}{\sqrt{6}}
\, \Big( \, d \big[ \frac{20}{3} D^3 +\frac{20}{3} D^2 F -8 DF^2 \big]
+ \, f \, \big[ \frac{4}{3} D^3 + 12 D^2 F - 8 DF^2 \big]\Big)\no \\
\beta_{\Xi^- \Lambda}^{(p)\,K} & = & 
-\frac{1}{m_\Xi-E_\Lambda} \frac{16}{\sqrt{6}} 
  \Big( -\frac{3}{2}h_3 - \frac{7}{2} h_5
              + \frac{1}{2} h_7 + h_8 \Big) (D-F)\no \\
& + &
\frac{1}{m_\Xi-m_\Sigma}  \frac{16}{\sqrt{6}}
                  \Big( -h_3 - h_5 - h_7 \Big) \, D \no \\
& + &
\frac{1}{\Lambda_\chi^2} \, \frac{1}{m_\Xi-E_\Lambda} \frac{1}{\sqrt{6}}
\,\Big( \, d \big[ -\frac{4}{3} D^3 - \frac{28}{3}D^2 F +4DF^2 -4 F^3 \big]
\no \\
&& \; + \, f \, \big[ -4 D^3 + 12 D^2 F + 12 D F^2 + 12 F^3 \big] \Big)\no \\
& + &
\frac{1}{\Lambda_\chi^2} \, \frac{1}{m_\Xi-m_\Sigma} \frac{1}{\sqrt{6}}
\, \Big( \, d \big[ \frac{16}{3} D^3 +\frac{8}{3} D^2 F + 8 DF^2 \big]
+ \, f \, \big[ 8 D^2 F + 8 D F^2\big]\Big)\no \\
\beta_{\Xi^- \Lambda}^{(p)\,\eta} & = & 
\frac{1}{\Lambda_\chi^2} \, \frac{1}{m_\Xi-E_\Lambda} \frac{2}{\sqrt{6}}
\,(d-3f) (D-F)(D+3F)F \no \\
& + &
\frac{1}{\Lambda_\chi^2} \, \frac{1}{m_\Xi-m_\Sigma} \frac{4}{3\sqrt{6}}
\, (d+f) D^2 (D+3F) \no \\
\gamma_{\Xi^- \Lambda}^{(p)\,\pi} & = & 
\frac{1}{m_\Xi-E_\Lambda}\frac{1}{\sqrt{6}} \frac{17}{12}  
                   \, (d-3f) (D-F)  +
\frac{1}{m_\Xi-m_\Sigma} \frac{1}{\sqrt{6}}\frac{17}{6}  D \, (d-f)
\no \\ 
& + &
\frac{1}{m_\Xi-E_\Lambda} \frac{1}{\sqrt{6}}
\, (D-F)^2 \,\Big( \, d (13D -4 F) - f \, ( 3 D - 12F) \Big) \no \\
& + &
\frac{1}{m_\Xi - m_\Sigma} \frac{1}{\sqrt{6}}
\, \Big( \, d \big[ \frac{22}{3} D^3 + 10 D^2 F + 4 DF^2\big]
+ \, f \, \big[ -\frac{2}{3} D^3 + 18 D^2 F + 4 DF^2\big]\Big)\no \\
\gamma_{\Xi^- \Lambda}^{(p)\,K} & = & 
 \frac{1}{m_\Xi-E_\Lambda} \frac{1}{\sqrt{6}}\frac{11}{6}  
                         (D-F) \, (d-3f) +
\frac{1}{m_\Xi - m_\Sigma}\frac{1}{\sqrt{6}} \frac{11}{3}  D \, (d+f)
\no \\ 
& + &
\frac{1}{m_\Xi-E_\Lambda} 
\,\Big(\, d \, \big[- \frac{2}{3} D^3 -\frac{38}{3} D^2 F +10 D F^2 
- 2  F^3\big]\Big) \no \\
&  &
+ f \, \big[ - 10 D^3 - 14 D^2 F + 6 D F^2 + 6 F^3\big]\Big) \no \\
& + &
\frac{1}{m_\Xi - m_\Sigma} \frac{1}{\sqrt{6}}
\,\Big(\, d \, \big[ 12 D^3 + 4 D^2 F + 8 D F^2 \big] 
+ f \, \big[ 4 D^3 +12 D^2 F + 8 D F^2\big]\Big) \no \\
\gamma_{\Xi^- \Lambda}^{(p)\,\eta} & = & 
\frac{1}{m_\Xi-E_\Lambda} \frac{1}{\sqrt{6}}\frac{3}{4}(D-F)\, (d-3f) +
\frac{1}{m_\Xi - m_\Sigma} \frac{1}{\sqrt{6}}\frac{3}{2}  D \, (d+f)
\no \\ 
& - &
\frac{1}{m_\Xi-E_\Lambda} \frac{1}{\sqrt{6}}
      ( d - 3f) (D+3F) (D-F)(\frac{1}{3}D - 2F) \no \\
& + &
\frac{1}{m_\Xi - m_\Sigma}\frac{1}{\sqrt{6}} (d+f)  2 D^2 
           \Big( \,\frac{7}{3}D  + 3 F\Big)\no \\
\epsilon_{\Xi^- \Lambda}^{(p)} & = &  
\frac{1}{\sqrt{6}} \Big( -12 g_{11} + 20 g_{13} + 4 g_{15} \Big) \no \\
\delta_{\Xi^- \Lambda}^{(p)} & = &  
-\frac{1}{m_\Xi-E_\Lambda} \frac{1}{\sqrt{6}}  (D-F) \, (d-3f) -
\frac{1}{m_\Xi - m_\Sigma}  2D \, (d+f)\no \\ 
\rho_{\Xi^- \Lambda}^{(p)} & = &
- \frac{1}{\sqrt{6}}
\frac{q^2 - (E_\Lambda - \mnod)^2}{(m_\Xi - E_\Lambda)^2}  (D-F)\, (d-3f)\no\\ 
\phi_{\Xi^- \Lambda}^{(p)} & = & - \frac{1}{\sqrt{6}} \, \Big( D-3F \Big)
\eeqa

\newpage

\section*{Table captions}

\begin{enumerate}

\item[Table 1] Experimental values of the decay amplitudes 
           including the errors.
           The numbers have to be multiplied by a factor of $10^{-7}$.

\item[Table 2] Numerical values of the LECs obtained from a 
           fit by using different values of the
           parameters $ F_{\pi}$, $ D$, 
           $F$, $ \mu $ and $ \mnod  $ and with the additional
           assumption $h_5 = h_7$.
           The first row shows the result for the central values 
           $ F_{\pi} = 93 $MeV, $ D= 0.75$, 
           $F=0.5$, $ \mu = 1.0$ GeV and $ \mnod = 767 $ MeV. 
           In the second row  $ \mnod = 940 $ MeV is used,
           $ D= 0.85$, $F=0.52$ in the third row.
           We changed the scale of dimensional regularization
           to $ \mu = 1.2$ GeV and $ \mu = 0.8$ GeV in the
           fourth and fifth row, respectively.
           The numbers have to be multiplied by a factor of $10^{-7}$.

\end{enumerate}

\vskip 1.2in

\section*{Figure captions}

\begin{enumerate}

\item[Fig.1] Baryon resonance excitation involving pion loops.
         The double line represents the decuplet. Solid and dashed
         lines represent the ground state octet baryons and the Goldstone
         boson fields, respectively. The solid square denotes
         $\Delta s =1$ weak interaction vertices and the solid dot
         vertices arising from the strong Lagrangian.

\item[Fig.2] Diagrams contributing to s-wave non-leptonic hyperon decays.
         Solid and dashed lines denote octet baryons and Goldstone bosons,
         respectively. The solid square represents a weak vertex and
         the solid circle denotes a strong vertex.

\item[Fig.3]  Diagrams contributing to p-wave non-leptonic hyperon decays.
         Solid and dashed lines denote octet baryons and Goldstone bosons,
         respectively. The solid square represents a weak vertex and
         the solid circle denotes a strong vertex.

\item[Fig.4]  Diagram with a weak decay of the meson. Solid and dashed 
         lines denote octet baryons and Goldstone bosons,
         respectively. The solid square represents a weak vertex and
         the solid circle denotes a strong vertex.

\end{enumerate}

\newpage

\begin{center}

\begin{table}[bht] 
\begin{center}
  \begin{tabular}{|cccc|}
    \hline
     ${\cal A}_{\Sigma^+ n}^{(s)}$ & 
      ${\cal A}_{\Sigma^- n}^{(s)}$ & 
      ${\cal A}_{\Lambda p}^{(s)}$ &  
      ${\cal A}_{\Xi^-\Lambda }^{(s)}$ 
        \\
    \hline
      0.13 $\pm$ 0.02 & 4.27 $\pm$ 0.02&  
          3.25 $\pm$ 0.02 &
         $-$4.51 $\pm$ 0.02  \\
      \hline
       \hline
     ${\cal A}_{\Sigma^+ n}^{(p)}$ & ${\cal A}_{\Sigma^- n}^{(p)}$ & 
      ${\cal A}_{\Lambda p}^{(p)}$ &  ${\cal A}_{\Xi^-\Lambda}^{(p)}$ \\
    \hline
      44.4 $\pm$ 0.16 &   $-$1.52 $\pm$ 0.16  & 23.4  $\pm$ 0.56
      & 14.8 $\pm$ 0.55 \\
      \hline
  \end{tabular}
  \medskip 
\end{center}
\end{table}
\vskip 0.7cm

Table  1

\vskip 1.5cm

\begin{table}[bht]\label{table2}
\begin{center}
  \begin{tabular}{|cccccccccc|}
    \hline
      $d$ & $f$ & $h_3$  
      & $h_5$ &  $h_7$ & $h_8$ & $g_{11}$ & $g_{13}$ & $g_{15}$ & $g_{16}$ \\
      $$[GeV] & [GeV]  
      & [GeV$^{-1}$] & [GeV$^{-1}$] & 
             [GeV$^{-1}$] & [GeV$^{-1}$] 
      & [GeV$^{0}$] & [GeV$^{0}$] & [GeV$^{0}$] & [GeV$^{0}$] \\
    \hline
 0.16 & $-$0.41 & 0.03 & 0.10 & 0.10& 0.10 &$-$0.48& $-$0.24 &0.44&$-$3.76 \\
 0.16 & $-$0.41 & 0.03 & 0.10 & 0.10& 0.10 &$-$0.47& $-$0.23 &0.44&$-$3.76 \\
 0.16 & $-$0.41 & 0.05 & 0.11 & 0.11& 0.13 &$-$0.47& $-$0.13 &0.46&$-$4.33 \\
 0.16 & $-$0.41 & 0.07 & 0.14 & 0.14& 0.19 &$-$0.45& $-$0.26 &0.39&$-$3.75 \\
 0.16 & $-$0.41 &$-$0.02 & 0.05 & 0.05&$-$0.02&$-$0.51& $-$0.21 &0.51&$-$3.78\\
     \hline
  \end{tabular}
  \medskip 
  \end{center}
\end{table}

\vskip 0.7cm

Table  2

\end{center}

\newpage

\begin{center}
 
\begin{figure}[bth]
\centering
\centerline{
\epsfbox{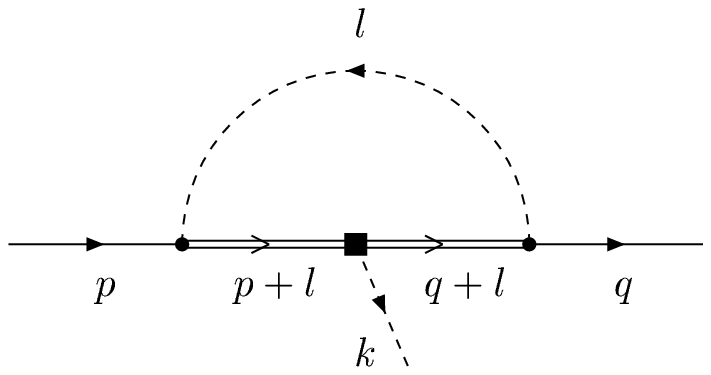}}
\end{figure}

\vskip 0.7cm

Figure 1

\vskip 1.5cm

\begin{figure}[tbh]
\centering
\centerline{
\epsfbox{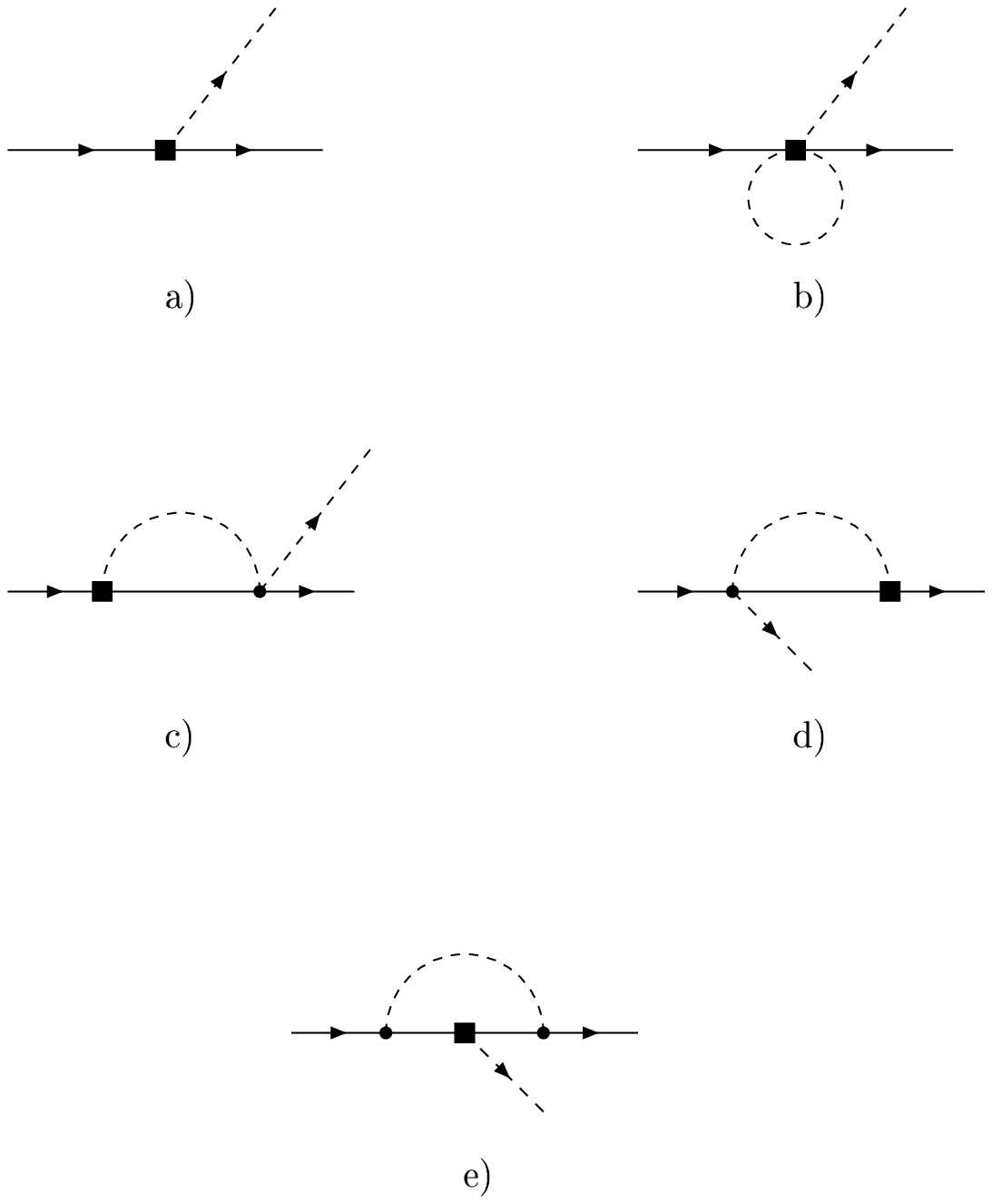}}
\end{figure}

\vskip 0.7cm

Figure 2

\vskip 1.5cm

\begin{figure}[tbh]
\centering
\centerline{
\epsfbox{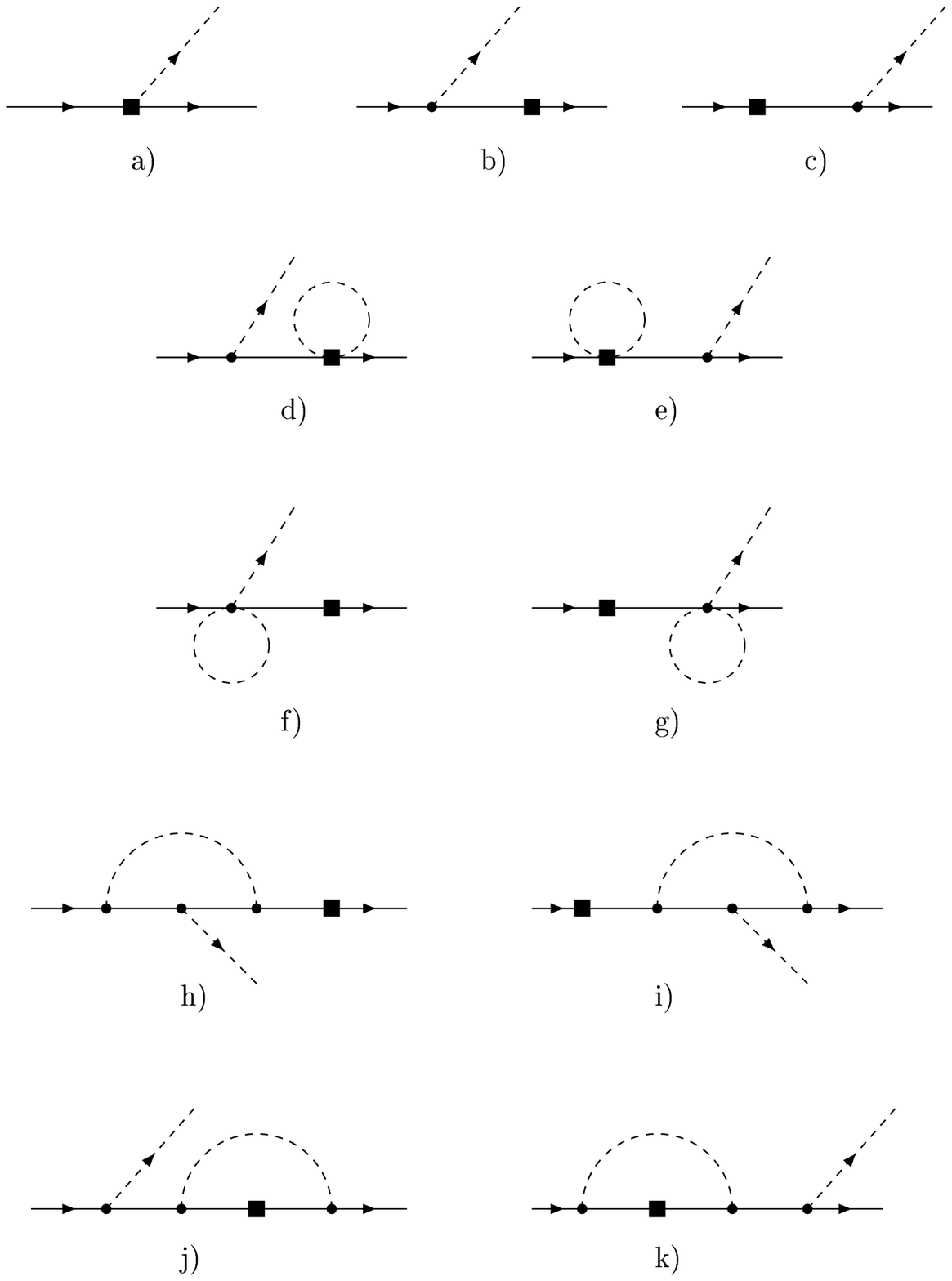}}
\end{figure}

\vskip 0.7cm

Figure 3

\vskip 1.5cm

\clearpage

\begin{figure}[tbh]
\centering
\centerline{
\epsfbox{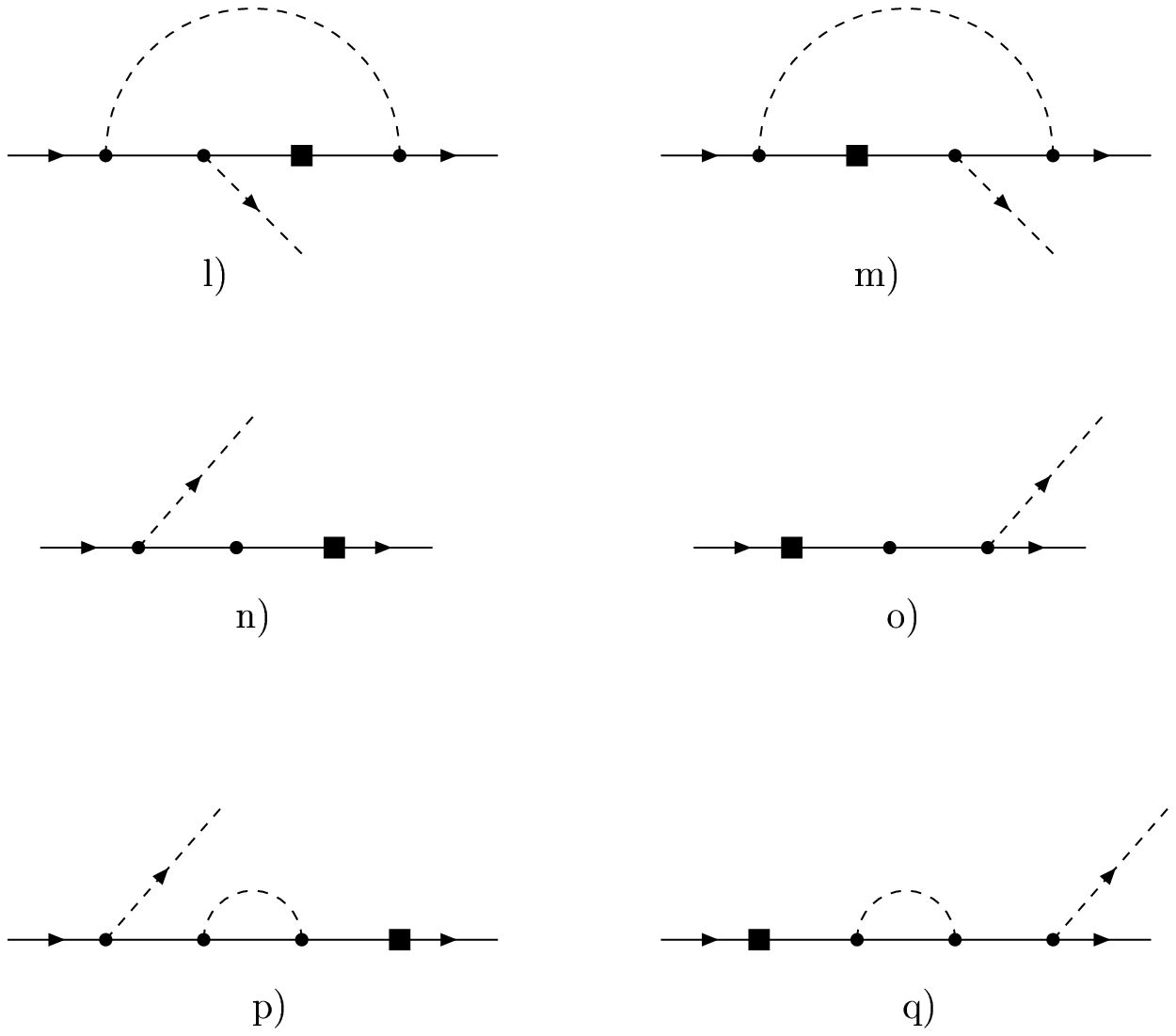}}
\end{figure}

\vskip 0.7cm

Figure 3 continued

\vskip 1.5cm

\begin{figure}[tbh]
\centering
\centerline{
\epsfbox{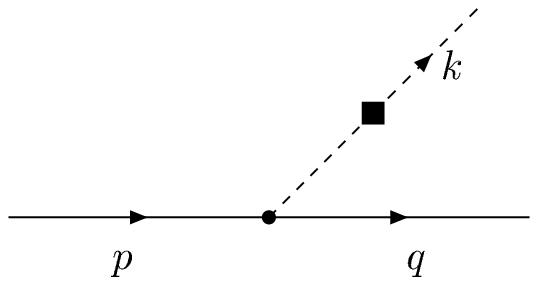}}
\end{figure}

\vskip 0.7cm

Figure 4

\end{center}

\end{document}